\documentclass[12pt]{JHEP3}
\def\a{\alpha}

\def\b{\beta}

\def\f{\frac}
\def\g{\gamma}

\def\l{\lambda}

\def\nn{\nonumber}
\def\o{\omega}

\newcommand{\be}{\begin{equation}}
\newcommand{\ee}{\end{equation}}
\newcommand{\bea}{\begin{eqnarray}}
\newcommand{\eea}{\end{eqnarray}}
\newcommand{\barr}{\begin{array}}
\newcommand{\earr}{\end{array}}

\title{Shear   sum rules at finite chemical potential}
\author{ Justin R. David {${}^a$}, Sachin Jain {${}^{b, c}$} and  Somyadip
Thakur {${}^a$}.  \\
${}^a$ Centre for High Energy Physics,
Indian Institute of Science,\\ C.V. Raman Avenue, Bangalore 560012, India. \\
\email{justin, somyadip@cts.iisc.ernet.in}\\
${}^b$Tata Institute for Fundamental Research \\
Homi Bhabha Road, Mumbai 400005, India.\\
${}^c$ Institute of Physics, 
Sachivalaya Marg,  \\ Bhubaneswar 751005, India. \\
\email{sachin@theory.tifr.res.in}\\ 
}

\abstract{ 
We derive sum rules which constrain the spectral density 
corresponding to the 
retarded propagator of the $T_{xy}$ component of the stress tensor
for three gravitational duals.  The shear sum rule is obtained for the 
gravitational dual of the ${\cal N}=4$ Yang-Mills,  theory of the M2-branes and M5-branes
all at finite chemical potential. 
We show that at finite chemical potential there are additional terms  
in the sum rule which involve the chemical potential. 
These modifications are shown to be due to the presence of  scalars 
 in the operator product expansion of the 
stress tensor which have non-trivial vacuum expectation values 
at finite chemical potential.  
}
\preprint{TIFR-TH/11-38}

\begin{document}

\section{Introduction}
Sum rules play an important role in constraining spectral densities of strongly coupled
fluids. There are important sum rules which constrain spectral densities of 
stress tensor in QCD  \cite{Baier:2009zy,Kharzeev:2007wb,Karsch:2007jc,Meyer:2010ii,Meyer:2010gu}
and the Ferrell-Glover-Tinkham sum rule satisfied  by
current-current correlation function  in the BCS superconductor 
\cite{Ferrell:1958zza,PhysRevLett.2.331}. 
The gauge gravity duality provides a framework for evaluating the 
two point functions of various conserved currents of strongly coupled
theories which admit a gravity dual. Thus one can use this framework 
to obtain sum rules.  Obtaining sum rules within this frame work provides information of the
analytic structure of the Green's functions in the strongly coupled limit. 
Romatschke and Son \cite{Romatschke:2009ng} derived two sum rules  for the 
stress tensor two point function for strongly coupled   ${\cal N}=4$ Yang-Mills theory.  
This was then generalized  for  Chamblin-Reall  backgrounds  which are dual to 
non-conformal theories in \cite{Springer:2010mf,Springer:2010mw}. 
The sum rule for the  R-charge correlator in strongly coupled 
${\cal N}=4$ Yang-Mills was  obtained in  \cite{Baier:2009zy}. 

As emphasized in \cite{Gulotta:2010cu},  sum rules are the consequences of analyticity of the 
Green's function in the complex frequency plane.  From the field theory point
of view this results from the  unitarity and causality of the field theory.  
 Deriving sum rules from gravity provides insight into 
how the properties of unitarity and causality of the boundary field theory are encoded in
the gravitational theory.  In \cite{Gulotta:2010cu} the differential equations which determine the
retarded Green's function of interest from gravity  were studied. The properties of these
differential equations were used  to obtain 
proof  and provide a unified framework to obtain various sum rules in 
gravity.  

In this paper we obtain the sum rules from gravity for the spectral density corresponding to the 
retarded propagator of the $T_{xy}$ component of the stress tensor.
This is done  for the case of  
${\cal N}=4$ Yang-Mills, the M2-brane and the M5-brane theory in the presence 
of chemical potential.  
One of our motivations 
to examine these situations is to determine how  the sum rules are modified when the 
dual theory  cannot be truncated to pure gravity.    The  presence of  chemical potentials 
in these systems results in additional scalars  in the gravitational theory. 
We will now briefly state the result for the ${\cal N}=4$ Yang-Mills case. 
The shear sum rule for ${\cal N}=4$ Yang-Mills derived in \cite{Romatschke:2009ng} 
in the absence of chemical potential is given by 
\begin{equation}
\frac{2}{5} \epsilon = \frac{1}{\pi} \int_{-\infty}^\infty  \frac{d\omega}{\omega}
\left( \rho(\omega) - \rho_{T=0} (\omega) \right),  
\end{equation}
where 
\begin{equation}
 \rho = {\rm Im } G_R(\omega), 
\end{equation}
and $G_R$ is the retarded propagator of the $T_{xy}$ component of the stress tensor. 
$\rho_{T=0}(\omega)$ is the spectral density at zero temperature and 
$\epsilon$ is the finite temperature energy density. 
In this paper we examine this sum rule in the presence of 
chemical potential.   We find that
the sum rule is modified to 
\begin{equation}
\label{modsum}
 \frac{2}{5} \epsilon  - \frac{N^2\pi^2 T_0^4}{120}
\left\{ ( k_1- k_2)^2 + ( k_1 - k_3)^2 + ( k_2-k_3)^2\right\} 
= \frac{1}{\pi} \int_{-\infty}^\infty  \frac{d\omega}{\omega}
\left( \rho(\omega) - \rho_{T=0} (\omega) \right).
\end{equation}
Where $k_i'$s  are functions of  the three chemical potentials  in 
${\cal N}=4$ Yang-Mills given in (\ref{cthermv}).
The relation between $T_0$  and the temperature of the 
Yang-Mills is given by  (\ref{cthermv}). 
Note that for the situation when all the charges or chemical potentials 
are equal then the correction vanishes.  For this case the the gravity background has 
no additional scalars present. 
We then show that these additional terms in the sum rule are 
due to the fact that, the 
operator product expansion of two stress tensor involves  operators of 
dimension 4 in addition to the stress tensor. 
We show that expectation values of these operators precisely have the 
same dependence in terms of the chemical potential to account for the 
additional terms.

For the M2-brane theory, the shear sum rule at finite chemical potential is given by 
\begin{eqnarray}
 & & \frac{3}{8} \epsilon +\f{\sqrt{2}\pi^2 N^{3/2} T_{0}^{3}}{216 } ( k_1 + k_2 - k_3-k_4) ( k_1 - k_2 + k_3 - k_4) ( k_1 -k_2-k_3+k_4) 
  \nonumber \\
& &  = \frac{1}{\pi} \int_{-\infty}^\infty  \frac{d\omega}{\omega}
\left( \tilde \rho(\omega) - \rho_{T=0} (\omega) \right). 
\end{eqnarray}
 Note that for the case of the M2-brane theory, there are 4-chemical potentials which
 are related to the  $k_i$'s by (\ref{chem4}) and $T_0$ is  related to the temperature by (\ref{temp4}). 
 $\tilde \rho$ is the imaginary part of the retarded Green's function shifted by a 
 term which is linear in frequency defined in (\ref{deltarho}). 
 The term proportional 
 to the energy density arises from the expectation value of the stress tensor in the OPE, 
 while the term proportional to the charge density arises from expectation value of 
 scalars. Finally for the M5-brane theory, the shear sum rule at chemical potential remains unchanged 
 in the presence of chemical potential. It is given by 
 \begin{equation}
  \frac{3}{7} \epsilon = \frac{1}{\pi} \int_{-\infty}^\infty  \frac{d\omega}{\omega}
\left( \rho(\omega) - \rho_{T=0} (\omega) \right). 
\end{equation}
In this case we  show the scalars which are turned on due to the chemical potential do 
not have the appropriate  conformal dimensions to occur in the OPE of the stress tensor and 
modify the sum rule.

The organization of the paper is as follows. 
In the next section we discuss the general considerations which go into deriving sum rules in 
quantum field theories. 
In  section 3.  we review the proof of the sum rule 
 given in \cite{Gulotta:2010cu} 
 for the case of ${\cal N}=4$ Yang-Mills in the absence of chemical potentials. 
 Here we develop  a  method to obtain the behaviour of the retarded Green's function 
at large frequencies using the Fefferman-Graham coordinates.  
This  method also allows us to extract the LHS of the sum 
rule easily. 
In section 4.  we derive the sum rule for D3-branes at finite chemical potential and explain 
the occurrence of additional terms in the sum rule (\ref{modsum}) due to the presence of 
expectation values of scalars. 
In section 5. we derive the sum rule for the M2 and M5-brane theories.
Section 6. contains our conclusions. 
 Appendix A contains the details of the Fefferman-Graham coordinates which are used to 
derive the sum rule.

\section{Sum rule generalities}

Consider the retarded Greens function  corresponding to an operator ${\cal O}$ of 
a quantum field theory in 4 dimensions given by 
\begin{equation}
\label{defgreen}
\tilde  G_R(t, x) \equiv i \theta(t) \langle [ {\cal O }( t, x), {\cal O}(0, 0)]\rangle , 
\end{equation}
we then take its Fourier transform given by 
\begin{equation}
\label{fttrans}
G_R(\omega, k) = \int d^4 x e^{i(\omega t - i k x)} \tilde G_R(t, x).  
\end{equation}
The spectral density  corresponding to this retarded correlator is defined as  
\begin{equation}
 \rho(\omega, k )  = \frac{1}{2i} \left\{ G_R(\omega, k ) - G_R(\omega, k)^* \right\}.  
\end{equation}
It is easy to see that  from (\ref{defgreen})  and from (\ref{fttrans}) we see that
for bosonic Hermitian operators we get
\begin{equation}
\label{realprop}
 G_R(\omega, k )^* = G_R(-\omega, -k). 
\end{equation}
In this paper we will restrict out attention to the case $k=0$ and we define 
$G_R(\omega)= G_R(\omega, 0).$ 
From the reality property (\ref{realprop}),  we see that in the Taylor series expansion of
$G_R(\omega)$  even powers of $\omega$ have real coefficients and odd powers of 
$\omega$ have purely imaginary coefficients. 
Sum rules for the spectral density are obtained   by assuming 
certain analytic properties of the  Green's function in the complex $\omega$-plane. \\
The  properties are the following
\begin{enumerate}
 \item $G_R(\omega)$ is holomorphic in the upper half plane, including the 
real axis. 
\item  $\lim_{|\omega| \rightarrow \infty} G_R(\omega) = 0$ if ${\rm{Im}}\,  \omega \geq 0$. 
\end{enumerate}
These properties will be referred to as {\emph{property 1}} and {\emph{ property 2}}
  in the rest of the paper. 
From the field theory point of view causality ensures  property 1 is satisfied.  
Using these properties,  we now indicate the arguments which go into deriving the sum rule.
By the first property  and by Cauchy's  theorem we have
\begin{eqnarray}
\label{cauchy}
 G_R(\omega + i\epsilon) &=&\frac{1}{2\pi  i} \oint_C \frac {G_R (z) dz}{ z -\omega -i\epsilon}, 
\\ \nonumber
0 &=& \frac{1}{2\pi  i} \oint_C  \frac {G_R (z) dz}{ z -\omega +i\epsilon}. 
\end{eqnarray}
for $\omega, \epsilon \in R$ and $\epsilon>0$. The contour $C$ is chosen 
such that it runs from $- r$ to $r$ for some large $r$ and then along the semi-circle
in the upper half plane and back to $-r$ with $r>0$. 
Because of the second property the integrals along the contour in (\ref{cauchy}) reduce to  
\begin{eqnarray}
\label{cauchy2}
G_R(\omega + i\epsilon) &=&  \frac{1}{2\pi i }\lim_{r\rightarrow\infty} \int_{-r}^r
\frac{G_R(z) dz}{ z-\omega - i \epsilon}, \\ \nonumber
0&=&    \frac{1}{2\pi i }\lim_{r\rightarrow\infty} \int_{-r}^r
\frac{G_R(z) dz}{ z-\omega + i \epsilon}. 
\end{eqnarray}
Then taking the difference of the two equations in (\ref{cauchy2})  we obtain
\begin{equation}
\label{cauchy3}
 G_R(\omega) = \lim_{\epsilon \rightarrow 0^+} \int_{-\infty}^\infty \frac{dz}{\pi} 
\frac{\rho(z)}{z -\omega - i\epsilon}, 
\end{equation}
where we have used the definition of the spectral density. 
Note that the integral runs  over the real line.  
The usual sum rule is obtained by evaluating the left hand side of 
(\ref{cauchy3})  at $\omega =0.$  Thus we  have
\begin{equation}
 G_R(0) = \lim_{\epsilon\rightarrow 0^+} \int_{-\infty}^\infty \frac{d\omega}{\pi} 
\frac{\rho(\omega)}{\omega - i\epsilon}. 
\end{equation}
In many of the situations  the Green's functions may not satisfy property 2, 
that is $\lim_{|\omega| \rightarrow \infty} G_R(\omega) = 0$ if ${\rm{Im}}\,  \omega \geq 0$.
In such situations  we  subtract divergences and obtain 
a regulated Green's function so that we can still ensure that property 
2  will be satisfied. 
The two point function of the stress tensor which we will study in this 
paper falls into this case. 

\section{The shear sum rule for uncharged D3-branes}

In this section we review the derivation of the shear sum rule  from gravity
for ${\cal N}=4$ Yang-Mills at zero chemical potential.  
This will help us set up notations and conventions.
To obtain the  retarded Green's function of the $T_{xy}$ component of the  stress tensor one 
needs to  solve the equation of the minimally coupled scalar field in the background of the 
D3-branes \cite{Policastro:2002se,Kovtun:2005ev}.  
Examining this differential equation it is possible to argue that
property 1 is satisfied.  To ensure property 2, 
we will develop a method to obtain the behaviour of the Green's function in gravity 
at large frequencies. This will enable us to regulate the Green's function
and ensure property 2,  which will lead us to the derivation of the sum rule.

\subsection{Green's function from gravity}

As we have mentioned earlier the  retarded  Green's function of interest in this paper is  
given by 
\begin{equation}
\label{greenret}
 G_R(t, \vec x) = i \theta( t) \langle [T_{xy}(t, \vec x) , T_{xy}(0, 0)] \rangle,
\end{equation}
where $T_{xy}$ is the $xy$ component of the stress tensor. 
The  dual geometry corresponding to ${\cal N}=4$ Yang-Mills at finite temperature is the
non-extremal D3-brane background.  The metric of this background  is given by 
\begin{eqnarray}
\label{d3back}
 ds^2 &=& \frac{r^2}{L^2} ( - f dt^2 + dx^2 + dy^2 + dz^2) + \frac{ L^2}{r^2 f} dr^2, 
\\ \nonumber
 f& =& 1 - \frac{r_+^4}{r^4} . 
\end{eqnarray}
The above metric is a solution of the action 
\begin{equation}
S= \frac{N^2}{8\pi^2 L^3} \int d^5 x \sqrt{g} \left( R + \frac{12}{L^2} \right).
\end{equation}
It is useful to recall the thermodynamic variables of this system.
The temperature, energy density, pressure and entropy density of this system are given by
\begin{eqnarray}
T= \frac{r_+}{\pi L^2},\qquad  \epsilon = \frac{3\pi^2 N^2 T^4}{8}, \\ \nonumber
P = \frac{\pi^2 N^2 T^4}{8}, \qquad s = \frac{\pi^2 N^2 T^3}{2}.
\end{eqnarray}
To evaluate the  Green's function given in (\ref{greenret}) 
from the gravity dual we first consider the metric  fluctuation which is 
dual to the stress tensor $T_{xy}$. This is given by 
\begin{equation}
\label{pertm}
 \delta g_{xy} = \phi(r) e^{-i\omega t + i kz} \frac{r^2}{L^2} .
\end{equation}
This perturbation obeys the equation of motion of a 
 minimally  coupled massless scalar  in the background (\ref{d3back}).   
The  $k=0$ mode satisfies the  following equation 
\begin{equation}
\label{mceq}
\partial_r^2 \phi +\left( \frac{F'}{F} + \frac{3}{r} \right) \partial_r \phi + 
\frac{\omega^2}{F^2} \phi =0,
\end{equation}
where
\begin{equation}
 F = \frac{r^2}{L^2} f.
\end{equation}
The procedure to obtain the Green's function \cite{Son:2002sd} is to first impose in going boundary conditions
at the horizon $r_+$ and obtain $\phi$ at the boundary $r\rightarrow\infty$. 
Once this is done, the Green's function at temperature $T$  is given by 
\begin{eqnarray}
\label{greendef}
 G_R(\omega, T) &=&  \hat G_R(\omega, T)  + G_{\rm{contact}}(T)  + G_{\rm{counter}}( \omega, T), \\ \nonumber
 \hat G_R(\omega) &=&  - \frac{N^2}{8\pi^2 L^6} \lim_{r\rightarrow \infty } \frac{ F r^3 \phi'}{\phi}, 
\end{eqnarray}
where $G_{\rm{contact}}(T)$ is the contribution from the contact terms obtained from the  on shell effective action. 
The on shell effective action is given by \cite{Gulotta:2010cu}. 
\begin{equation}
\label{os}
 S_{\rm{os}} = P \int d^4 x \left. \left( 1 - \frac{1}{2} \phi^2 \right)\right|_{r\rightarrow\infty} ,
\end{equation}
where $P$ is the pressure of the solution. 
Evaluating  $G_{\rm{contact}}$  we obtain
\begin{equation}
 G_{\rm{contact}} =-P.
\end{equation}
$G_{\rm{counter}}( \omega, T)$  is the contribution from the counter terms required to cancel
the $r^2$ and $\log (r)$ divergences in $G_R(\omega)$. The details of these terms are given in 
$\cite{Gulotta:2010cu}$, but as we will  see  later, we will not need the detailed structure of these terms. 
An important point to note is that  the on-shell action is independent  of frequency
which ensures that the contact term $G_{\rm{contact}}$  is also independent 
of frequency. 

Thus the important  properties of the Green's function is essentially contained 
in the function
\begin{equation}
\label{defgr}
 g_R(\omega) =  \lim_{r\rightarrow \infty } \frac{ F r^3 \phi'}{\phi}.
\end{equation}
Therefore  to study the behaviour of the retarded Green's function, 
it is sufficient to study the function  $g_R(\omega)$ whose behaviour can be 
understood by examining the equation (\ref{mceq}).

Before we go ahead, we first discuss the asymptotic properties of the 
solution of the differential equation (\ref{mceq}). 
Near the horizon the solutions are wave like and is given by 
\begin{equation}
 \phi (r) \sim ( r- r_+)^{ \pm \frac{i \omega}{F_h} }, \qquad
r\rightarrow r_+, 
\end{equation}
where $F_h$ is given by the expansion 
\begin{equation}
 F = ( r-r_+) F_h + \cdots, \qquad F_h = 4 \frac{r_+}{L^2}.
\end{equation}
Now at the boundary the two independent solutions are given by
\begin{eqnarray}
\label{bexp}
 \phi(r) &\rightarrow& \frac{L^4\omega^2}{r^2} J_2( \frac{L^2\omega}{r} ) \sim r^{-4},
 \quad r\rightarrow\infty,  \\ \nonumber
 \phi(r) &\rightarrow&  \frac{L^4\omega^2}{r^2} K_2( i \frac{L^2\omega}{r} ) \sim 
 {\rm{constant}},  \quad
 r\rightarrow \infty.
\end{eqnarray}
We will also need the fact that the differential equation given in (\ref{mceq})
can be obtained as the equations of motion of the following  action
\begin{equation}
\label{varact}
S_\phi  = \int _{r_h}^\infty dr Fr^3 \left( 
|\phi'(r)|^2 - \frac{\omega^2}{F^2} |\phi(r)|^2 \right) .
\end{equation}

\subsection{Green's function in the $\omega$-plane} 

In this subsection we will discuss the analytic properties of the function $g_R(\omega)$ 
defined in (\ref{defgr}) in 
the complex $\omega$-plane.  

\vspace{.5cm}
\noindent
{\bf No poles  for ${\rm{Im}} \, \omega >0$.} 
\vspace{.5cm}

Poles or divergences $g_R(\omega)$ correspond to quasi-normal modes of the equation
(\ref{mceq}) \cite{Kovtun:2005ev}. Quasi-normal modes are   solutions to the differential 
equation (\ref{mceq})  with the following boundary conditions. 
\begin{eqnarray}
\label{bccond}
 \phi( r) &\sim& ( r - r_+) ^{ -i\frac{\omega}{F_h}}, \qquad r\rightarrow r_+, \\ \nonumber
\phi(r) &\sim& r^{-4} , \qquad r \rightarrow \infty.  
\end{eqnarray}
We will now prove that such quasi-normal modes do not 
exist. A simple intuitive reason that prohibits such quasi-normal modes
is that,  if quasi-normal modes with 
${\rm{Im}} \, \omega >0$ exists,  they will correspond to instabilities due to  the time
dependence $\exp(-  i\omega t)$ in the mode given in (\ref{pertm}). 
We will now demonstrate such quasi-normal modes do 
not exist directly from the equation (\ref{mceq}).  
All the coefficients in the equation  (\ref{mceq})  are real. Therefore
if $\phi(r)$ is a quasi-normal mode with 
say complex frequency $\omega$, then $\phi(r)^*$ the complex conjugate is 
also a quasi-normal mode with frequency $\omega^*$. 
This is seen by just taking the complex conjugate of the 
equation (\ref{mceq}). 
We can now consider $S_\phi- S_\phi =0$ which by integration by parts and using 
equations of motion we obtain
\begin{equation}
 0 = Fr^3 ( \phi^{*\prime} \phi - \phi^* \phi') |_{r_h}^\infty
+ ( \omega^{*2} - \omega^2 ) \int_{r_h}^\infty dr \frac{ r^3}{F} |\phi|^2.
\end{equation}
Here we have used the equations of motion of $\phi^*$ in the first $S_\phi$ and 
the equation of motion of $\phi$ in the second $S_\phi$. 
Now from the conditions for the quasi-normal modes given in 
(\ref{bccond}) and the fact that $F\sim r^2$ as $r\rightarrow \infty$
and $ F \sim ( r- r_h) F_h$ as $ r\rightarrow r_h$ 
we see that the first term in the above equation vanishes 
for ${\rm{Im}}( \omega ) >0$. Then we have the case that 
\begin{equation}
 \omega^2 =  \omega^{*2},
\end{equation}
since the second term is positive definite.
Thus we see that $\omega$ has to be either purely real or purely imaginary, 
now with the condition ${\rm{Im}}( \omega ) >0$ we find that
$\omega$ is restricted to be on the upper imaginary axis. 

Let us now assume that there exists a quasi normal modes on the upper
imaginary axis with $\omega^2 <0$. Then $S_\phi$ is positive definite. 
Substituting this solution in $S_\phi$ and  integrating by parts we obtain 
\begin{equation}
 S_\phi = F r^3 \phi^*(r) \phi'(r) |_{r_h}^\infty.
\end{equation}
This vanishes from the behaviour in (\ref{bccond}) and the fact that 
 ${\rm{Im}}\,  \omega  >0$. But since $S_\phi$ is positive definite 
we must have $\phi=0$. 
Thus no quasi-normal modes exist in the upper half plane which 
implies no poles or divergences exist in the Green's function in this domain.

\vspace{.5cm}
\noindent
{\bf No poles for  $\omega$ real and  $\omega \neq 0$}
\vspace{.5cm}

For real $\omega,$   the following quantity 
\begin{equation}
\label{defwronk}
 W = \phi'  \phi^* -  \phi^{\prime *} \phi,
\end{equation}
is the Wronskian. From the differential equation in 
(\ref{mceq}) we see that the Wronskian satisfies the equation
\begin{equation}
W' + \left( \frac{F'}{F} + \frac{3}{r} \right) W =0.
\end{equation}
The solution of this equation is given by
\begin{equation}
W = \frac{C}{r^3 F}.
\end{equation}
We can determine the  constant $C$ since we know that if there exists a quasi-normal mode, 
the behaviour of the mode near
 the horizon  is given by the ingoing boundary conditions given in 
(\ref{bccond}). This fixes the $C$ which determines the Wronskian to be  given by 
\begin{equation}
W = -i\frac{2\omega r_+^3}{ r^3 F}.
\end{equation}
From here we see that the quantity  $r^3  F W$ does not vanish any where. 
 We can now examine its behaviour as $r\rightarrow\infty$ for the quasi-normal mode. 
 The  boundary conditions for the quasi-normal mode given in (\ref{bccond}) implies that 
 $\phi \rightarrow r^{-4}$ as $r\rightarrow\infty$.   Using this fact in the definition 
 given in (\ref{defwronk})  we
 conclude the quantity $r^3 FW$ must vanish as
$r \rightarrow \infty$ which contradicts our previous conclusion. Thus quasi-normal modes
and hence poles or divergences of the Greens's function do not exist on the real line
for $\omega \neq 0$. 

\vspace{.5cm}
\noindent
{\bf  No poles for   $\omega =0 $}
\vspace{.5cm}

We can determine the behaviour of the function $g_R$ as $\omega\rightarrow 0$ 
explicitly from the differential equation (\ref{mceq}) as follows. 
Let us define 
\begin{equation}
\tilde g(r)  = \f{\phi'(r)}{\o \phi(r)}.
\end{equation}
We can use the differential equation (\ref{mceq}) to obtain the equation satisfied by 
$\tilde g$. This is given by 
\begin{equation}
\label{omg0eq}
\tilde g '(r)+\o \tilde g ^{2}(r)+\left(  \f{F'}{F}+\f{3}{r}\right) \tilde g (r)+ \f{\o}{F^2}=0.
\end{equation}
The ingoing boundary conditions at the horizon given in (\ref{bccond}) determines
the boundary condition for $\tilde g,$ this is given by 
\begin{equation}
\label{bccond1}
\tilde g \sim \f{-i}{F_{h}(r-r_+)}, \qquad r\rightarrow r _+ .
\end{equation}
The equation (\ref{omg0eq}) can be easily solved in the $\omega\rightarrow 0$ limit. 
The solution which agrees with the  boundary condition in (\ref{bccond1}) is given by 
\begin{equation}
\tilde g = -i \frac{r_+^3}{r^3 F}, \qquad \omega\rightarrow 0.
\end{equation}
This implies 
\begin{equation}
\label{leador}
g_R = \omega\lim_{r\rightarrow \infty} r^3 F \tilde g 
= -i\omega  r_+^3, \qquad \omega\rightarrow 0.
\end{equation}
Thus we conclude that the Green's function has no poles as $\omega\rightarrow 0$. In fact 
as can be seen from the equation (\ref{omg0eq}), 
it admits an analytic power series  expansion in $\omega$ at the origin. The leading behaviour
is given above in (\ref{leador}). 
Using  (\ref{omg0eq}) one can easily set up a
perturbative expansion in $\omega$ and solve for $g_R(\omega)$ order by order in $\omega$ which 
gives rise to the power series expansion around the origin.

\vspace{.5cm}
\noindent
{\bf No zero's for ${\rm Im}\, \omega \geq 0, \omega \neq 0$}
\vspace{.5cm}

Just as quasi-normal modes correspond to Dirichlet boundary conditions for the 
field $\phi$ at the boundary which leads to the boundary conditions  given in (\ref{bccond})
zeros of the Green's function correspond to Neumann boundary conditions for the 
field $\phi$. This can be seen from the definition (\ref{defgr}) which implies that 
zeros of the Green's function correspond to the condition 
\begin{eqnarray}
\label{bccondn}
\phi( r) &\sim& ( r - r_+) ^{ -i\frac{\omega}{F_h}}, \qquad r\rightarrow r_+, \\ \nonumber
r^3 F \phi'(r) &\rightarrow& 0, \qquad r\rightarrow\infty.
\end{eqnarray} 
One might suspect that there might be singularities in $\phi$ which also gives 
zeros of the Green's function. However as discussed in \cite{Kovtun:2005ev}, a theorem of 
\cite{Arnold} guarantees that the solution is smooth with respect to a parameter $\omega$ if the 
differential equation and the boundary conditions are smooth with respect to it. 
This ensures that there cannot be singularities in $\phi$ in the $\omega$ plane. 
Now with the Neumann boundary conditions given in (\ref{bccondn}) we can repeat the 
steps of the discussion for the absence of poles or singularities in the Green's function. 
All the steps goes through since, here again
the boundary conditions in (\ref{bccondn}) ensures that 
\begin{equation}
Fr^3 ( \phi^{*\prime}\phi - \phi^*\phi') =0, \qquad r\rightarrow r_+, r\rightarrow \infty.
\end{equation}
Similarly for ${\rm Im}\, \omega =0, \omega\neq 0$ we have the limit
\begin{equation}
WFr^3 \rightarrow 0, \qquad r \rightarrow\infty,
\end{equation}
where $W$ is the Wronskian. 
These conditions allows us to conclude that 
the Green's function does not vanish in the upper half $\omega$ plane including the 
real axis, with the exception of $\omega=0$ where we have shown that it vanishes and it 
admits an analytic power series expansion.

\vspace{.5cm}\noindent
{\bf Absence of branch cuts for ${\rm Im }\, \omega \geq 0$\footnote{We would like to thank Ashoke Sen and the anonymous 
referee of this paper for useful comments on this section.}}
\vspace{.5cm}

Recall that the 
important properties of the Green's function is essentially contained in the 
function $g_R(\omega)$ as defined in (\ref{defgr}).
In this equation $\phi (r)$ is 
the solution for the minimally coupled scalar equation in the background of 
(\ref{d3back}) with ingoing boundary (\ref{nrhor}) at the 
horizon. From (\ref{defgr}) we see that that we need to examine the 
behaviour of solution $\phi(r)$ and its derivative at the boundary. 
Note that the 
differential equation for the minimally coupled scalar and the boundary condition depends on the parameter $\omega$ and both 
these dependences are smooth with respect to this parameter $\omega$. 
Now the  general theorem in \cite{Arnold} {\footnote{
In fact this theorem has been used in the identification of the location of the 
poles of the Green's function with 
the quasi-normal modes of  $\phi$  in \cite{Kovtun:2005ev}. 
Since in general the poles of the Green's function can also arise from the 
singularities of $\phi'$.  However this theorem guarantees that 
$\phi'$ is smooth with respect to $\omega$ hence poles can arise only 
from the zeros of $\phi$ which correspond to its quasi-normal modes. }
ensures that the local Forbenius expansion of the solution is smooth with respect to 
the parameters of the differential equation if both the equation and the boundary condition are smooth with respect to the parameter. Hence 
the Forbenius expansion of $\phi$ and its radial derivative 
at the boundary  are smooth with respect to the parameter $\omega$.
Now using this theorem it is easy to see that the $n^{\rm th}$ order derivative of the $g_R(\omega)$ 
with respect to $\omega$
is smooth  in the upper half $\omega$ plane. 
Since $\phi$ and its derivative are smooth with respect to $\omega$ we see from 
 (\ref{defgr}) the only possible locations of singularities 
  of the $n^{\rm th}$ order derivative of the $g_R(\omega)$ 
  are the zeros of $\phi$, that is where it satisfies the quasi-normal mode  boundary conditions
 given  in (\ref{bccond}). 
  However we have shown earlier that there 
  are no-quasi-normal modes  in the upper half-plane,  this assures that 
   arbitrary derivatives of $g_R(\omega)$ exists  in the upper half plane. 
  This implies that the Green's function does not have any  branch cuts in the upper half-plane.

Thus we have concluded that the Green's function satisfies property 1 for the derivation of the
sum rule. We now need to ensure that property 2 is satisfied.

\vspace{.5cm}\noindent
{\bf  Behaviour as $\omega \rightarrow \infty$}
\vspace{.5cm}

To obtain the behaviour of the Green's function at large $\omega$ we first 
rewrite the differential equation given in (\ref{mceq}) by defining 
a dimensionless variables 
\begin{equation}
 z = \frac{r_+}{r},  \quad  i  \lambda =  \frac{L^2}{r_+} \omega . 
\end{equation}
The equation reduces to  
\bea
\label{largom}
\phi''(z)-\f{1}{z f(z)}(3+{z^4})\phi'(z)-\f{\l^2}{f(z)^2}\phi(z)=0,
\eea
where $f(z)=1-z^4$. 
Note that for convenience we have gone over to the Euclidean frequency labelled by $\lambda.$ 
Let us re-state the boundary conditions in terms of these coordinates. 
The ingoing boundary condition reduces to 
\begin{equation}
 \phi( z) \sim ( 1-z)^{ \frac{\lambda}{4} } , \qquad z \rightarrow 1 
\end{equation}
 and the Green's function is given by 
\begin{equation}
 g_R(\lambda) =  - \frac{r_+^4}{L^2} \lim_{z\rightarrow 0}  \frac{\phi'}{z^3 \phi}.
\end{equation}
Our aim now is to find  the behaviour of this function as $\lambda \rightarrow\infty.$ 
For this purpose we first rescale the co-ordinates as 
\begin{equation}
 y = \lambda z.
\end{equation}
The differential equation  given in (\ref{largom}) can be written as 
\begin{equation}
\label{largom1}
 \phi''(y)-\f{1}{y f(y)}(3+\f{y^4}{\lambda^4} )\phi'(y)-\f{1}{f(y)^2}\phi(y)=0,
\end{equation}
where
$f  =  1- \frac{y^4}{\lambda^4}$ and derivatives are with respect to $y.$
We can solve   the equation using the Forbenius
expansion at around $y= \lambda.$ This  results in  following series
\begin{eqnarray}
\label{horexp} 
\phi &=& ( 1-\frac{y}{\lambda} ) ^{\frac{\lambda}{4}} 
\left( \sum_{j=0}^\infty a_j ( 1- \frac{y}{\lambda})^j \right).
\end{eqnarray}
The coefficients $a_j$ can be determined by recursion relations on substituting the Forbenius 
expansion in the differential equation (\ref{largom1}).  
Now the differential equation in (\ref{largom1}) also admits an expansion at around $y \rightarrow 0.$ 
This expansion can be organized as a systematic expansion in powers of $\frac{1}{\lambda}$ as follows:
We first expand (\ref{largom1}) in powers of $1/\lambda$, this results in 
\be
\label{largomexp}
\phi''(y)-\f{1}{y}(3+4\f{y^4}{\l^4}+ 4\f{y^8}{\l^{8}}
+\cdots )\phi'(y)-(1+2\f{y^4}{\l^4}+3\f{y^8}{\l^8} + \cdots)\phi(y)=0.
\ee
One important fact to point out in this expansion is that at $\lambda$ strictly infinity, the 
equation reduces to that of the minimally coupled equation in the background of pure $AdS_5$, that is 
the background corresponding to  zero temperature. 
To solve this equation perturbatively in $\frac{1}{\lambda}$ we define the quantity
\begin{equation}
 g(y) = \frac{\phi'(y)}{\phi(y)}.
\end{equation}
From the expansion of the equation in (\ref{largomexp}),  we see that the we can expand $g$ 
as 
\begin{eqnarray}
 g &=& g_0 (y) + \frac{1}{\lambda^4} g_1(y)  + \frac{1}{\lambda^8} g_2(y)  + \cdots  \\ \nonumber
&=& \sum_{n=0}^\infty \frac{1}{\lambda^{4n}} g_n.
\end{eqnarray}
We can obtain the equations satisfied by each $g_n$ by substituting the expansion above 
and matching
orders in $\frac{1}{\lambda^4}.$ The first few equations are given by 
\begin{eqnarray}
\label{pereqn}
& & g_0'(y)+g_{0}^{2}(y)-\f{3}{y}g_0(y)-1=0,\nn\\
& & g_1'(y)+\left(  2g_0(y)-\f{3}{y}\right)  g_1(y)-4y^3 g_0(y)-2y^4=0,\\ \nonumber
& & g_2'(y)+\left(  2g_0(y)-\f{3}{y}\right)  g_2(y)+g_{1}^{2}(y)-4y^3 g_1(y)-4y^7g_0(y)-3y^8=0. 
\end{eqnarray}
The advantage of casting the equations in this form is that the apart form the first equation, the subsequent  equations  are first order equations. 
We will now illustrate how to obtain the two independent solutions. There are two solutions 
to the first equation, these are given by 
\begin{eqnarray}
\label{zorder}
 g_0^{(1)} = -\f{K_1(y)}{K_2(y)}= \f{d}{dy}\left( \log(y^{2}K_{2}(y))\right), \;\;
g_0^{(2)} = \f{I_1(y)}{I_2(y)}= \f{d}{dy}\left(  \log(y^{2}I_{2}(y))\right).
\end{eqnarray}
Let us first find the $\frac{1}{\lambda}$ expansion around the first solution, for that we have to solve
for $g_1$, this is given by 
\begin{eqnarray}
 g_1^{(1)}(y)&=&\f{1}{y K_{2}^{2}(y)}\int_0^y dy[4y^3 g_0^{(1)}(y)+2y^4](y K_{2}^{2}(y)) 
+ \frac{c_1}{y K_{2}^{2}(y)}, \nn\\
&=&2y^3+\f{y^5}{5}(1-\f{K_{3}^{2}}{K_{2}^{2}}) +\frac{c_1}{ y K_{2}^{2}(y)}.
\end{eqnarray}
The term proportional to $c_1$ is the homogeneous solution. 
The constant $c_1$ is set to zero by requiring  that the asymptotics of the solution set by $g_0$ 
 be unchanged in the presence of $g_1$.  Note that if $c_1\neq0$, $g_1$ grows exponentially 
when $y\rightarrow\infty.$ This condition is imposed on all equations.  Thus we have
\begin{eqnarray}
\label{og11}
 g_1^{(1)}(y)&=&2y^3+\f{y^5}{5}(1-\f{K_{3}^{2}}{K_{2}^{2}}),  \\ \nonumber
&=& - \frac{6}{5} y^3 - \frac{3}{5} y^5 + O ( y^7,y^7 \log(y)). 
\end{eqnarray}
Similarly the solution of $g_2$ is given by 
\begin{equation}
g_2(y)=\f{1}{y K_{2}^{2}(y)}\int_0^y
 dy[-( g_{1}^{(1)}) ^{2}(y)+4y^3 g_1^{(1)} (y)+4y^7g_0^{(1)}(y)+3y^8](y K_{2}^{2}(y)).
\end{equation}
From the expansion of the functions involved, it can be shown  that $g_2^{(1)}$ admits the 
following expansion near the origin
\begin{equation}
 g_2^{(1)}(y)\sim -\f{156}{25}y^7+y^9+(\log(y)+\log(2)+\g) y^{11} + \cdots ~.
\end{equation}
An simple examination of the equation for $g_n$ shows that the 
leading term in the expansion of $g_n^{(1)}$ around the origin is given by 
\begin{equation}
\label{og1n}
 g_n^{(1)} (y) \sim y^m , \quad m \geq 7, \quad \hbox{for} \; n\geq 2 .
\end{equation}
Now that we have $g_n$, the 
first solution $\phi^{(1)} (y))$ is given by 
\begin{eqnarray}
\label{lamexp}
 \phi^{(1)}  (y) &=& \exp( \int_0^y dy  \sum_{n=0}^\infty \frac{g_n}{\lambda^{4n}} ),  \\ \nonumber
&=&  y^2 K_2(y) \left( 1 + \frac{1}{\lambda^4 } \int_0^y dy g_1^{(1)} ( y)   \right. \\ 
\nonumber & &  + \frac{1}{\lambda^8}
 \left.  \frac{1}{2} \left[  \int_0^y dy g_1^{(1)} ( y) )^2 +   \int_0^y g_2^{(1)} (y)  \right]  + \cdots \right) .
\end{eqnarray}
Note that $\phi^{(1)}$ has the following behaviour as $y \rightarrow 0$
\begin{equation}
 \phi^{(1)} \sim {\rm{constant}}, \qquad y\rightarrow 0.
\end{equation}
From (\ref{og11}) we see that the $\frac{1}{\lambda^4}$ coefficient in the 
expansion given in (\ref{lamexp})  behaves as $y^4$ near the origin. 
 The equation ( \ref{og1n}) shows that the remaining terms are further 
suppressed as $y\rightarrow 0$. 

A similar construction can be done starting with the seed
$g_0^{(2)}$. 
The first $\f{1}{\l^{4}}$ expansion around the second  solution is given by 
\begin{eqnarray}
\label{og21}
 g_1^{(2)}(y) &=&
 \f{1}{y I_{2}^{2}(y)}\int_0^y dy[4y^3 g_0^{(2)} (y)+2y^4](y I_{2}^{2}(y)),  \nn\\
g_1^{(2)} (y) &=&2y^3+\f{y^5}{5}(1-\f{I_{3}^{2}}{I_{2}^{2}}). 
\end{eqnarray}
Note that we have set the constant corresponding to the homogeneous solution to zero. 
Similarly the solution of $g_{2}^{(2)}$ is given by 
\begin{equation}
g_2^{(2)} (y)=\f{1}{y I_{2}^{2}(y)}\int_0^y
 dy[-( g_{1}^{(2)})^{2}(y)+4y^3 g_1^{(2)} (y)+4y^7g_0^{(2)} (y)+3y^8](y I_{2}^{2}(y)).
\end{equation}
From the expansions of the functions involved it can be shown that 
\begin{equation}
g_2^{(2)}(y)  \sim y^7, \quad y \rightarrow 0.
\end{equation}
Similarly in general it can be shown that 
\begin{equation}
\label{og2n}
g_n^{(2)}(y)  \sim y^m, \quad m \geq 7, \quad \hbox{for}\; n \geq 2, \quad y \rightarrow 0.
\end{equation}
Thus we obtain the second solution
\begin{eqnarray}
\label{lamexp1}
 \phi^{(2)} (y) &=& y^2 I_2(y) \left( 1 + \frac{1}{\lambda^4 } \int_0^y dy g_1^{(2)} ( y)   \right. \\ 
\nonumber & &  + \frac{1}{\lambda^8}
 \left.  \frac{1}{2} \left[  \int_0^y dy g_1^{(2)} ( y) )^2 +   \int_0^y g_2^{(2)} (y)  \right]  + \cdots \right) .
\end{eqnarray}
It will be useful to note that $\phi^{(2)}$ has the following behaviour as $y\rightarrow 0$. 
\begin{equation}
 \phi^{(2)} \sim y^4, \qquad y \rightarrow 0. 
\end{equation}
Note that from (\ref{og21}) and (\ref{og2n}) we see that the coefficients of the $\frac{1}{\lambda}$
expansion are further suppressed as $y\rightarrow 0$.

Once  the solutions $\phi^{(1)}, \phi^{(2)}$  have been constructed, then the 
solution which matches the ingoing boundary condition at the horizon is obtained
as follows: Considering the linear combination
\begin{equation}
\label{matsol}
 \phi(y) = A(\lambda) \phi^{(1)} (y) + B(\lambda)  \phi^{(2)}(y) .
\end{equation}
The coefficients $A$, $B$  which depend on $\lambda$ are  obtained by 
matching the solution $\phi(y)$  in (\ref{matsol})  and that obtained by the Forbenious 
expansion at $y =\lambda$ given in (\ref{horexp}).
The expansion in (\ref{horexp}) is asymptotically expanded  to $y\rightarrow 0$, while 
the $\frac{1}{\lambda}$ expansion in (\ref{lamexp}) is expanded to $y\rightarrow \infty$ and the 
function and the derivatives are matched to determine $A$ and $B$. 
Though we do not explicitly require $A$ and $B$ as functions of $\lambda$,  we can 
argue that 
\begin{equation}
\label{basym} 
B(\lambda)  \rightarrow 0, \qquad 
A(\lambda) \rightarrow \frac{1}{2}, \qquad y \rightarrow \infty.
\end{equation}
The reason for this is that 
when $\lambda$ strictly $\infty$, the equation (\ref{largomexp}) reduces to that 
of the zero temperature case  as we have observed earlier. In  this situation
the solution which is finite at $y\rightarrow 0$  is given by $\frac{1}{2}y^2 K_2(y)$
\footnote{The $\frac{1}{2}$ is a normalization such that the solution  $\frac{1}{2}y^2 K_2(y) \rightarrow 1$ as $y\rightarrow 0$. }. Thus we arrive at (\ref{basym}). 

Now let us study the implications of the properties of  the solutions of the differential equation 
(\ref{largom1})  have on the retarded Green's function $g_R(\omega)$. In terms of the 
scaled variables it is given by  
\begin{equation}
\label{grelim}
 g_R(\lambda) = - \frac{r_+^4}{L^2}  \lim_{y\rightarrow 0} \frac{\lambda^4}{y^3} 
\frac{ A(\lambda) \phi^{(1)'} (y) +  B(\lambda)  \phi^{(2)'}( y) }
{A(\lambda) \phi^{(1)} (y) +  B(\lambda)  \phi^{(2)}( y) }. 
\end{equation}
 Now from the behaviour  of the solutions
as $y\rightarrow 0$ given in  (\ref{og11}), ( \ref{og1n}), (\ref{og21}), (\ref{og2n})  and 
the discussions below (\ref{lamexp}) , (\ref{lamexp1})  
we see that the $y\rightarrow 0$ limit in (\ref{grelim}) reduces to 
\begin{eqnarray}
 g_R(\lambda) &=&  - \frac{r_+^4}{L^2}\lim_{y\rightarrow 0} \left( 
\frac{1}{y^3} \left( \lambda^4 g_0^{(1)} (y)   + g_1^{(1)} \right) 
+ \frac{1}{4} \frac{B (\lambda) }{A(\lambda)}   \right) \\ \nonumber
&=&   - \frac{r_+^4}{L^2}\left(  \lim_{y\rightarrow 0} 
\frac{1}{y^3}  \lambda^4 g_0^{(1)} (y)   - \frac{6}{5}  + 
\frac{1}{4} \frac{B (\lambda) }{A(\lambda)}  \right). 
\end{eqnarray}
In the second line  of the 
above equations we have substituted the expansion of  $g_1^{(1)}(y)$ near the origin which is 
given in (\ref{og11}).   
An important point to note is that the above result is true for all values of $\lambda$ or 
frequency. 
We can  further take the limit $\lambda\rightarrow \infty$  and using 
(\ref{basym})  we are left with 
\begin{eqnarray}
 \lim_{\lambda\rightarrow \infty} g_R(\lambda)  &=&  - \frac{r_+^4}{L^2} \left( \lim_{y\rightarrow 0}
 \frac{1}{y^3}  \lambda^4 g_0^{(1)}    - \frac{6}{5}  \right) .
\end{eqnarray}
As we have observed the leading  contribution $g_0^{(1)}$ is identical to the zero temperature case, 
it is divergent as $\lambda\rightarrow\infty$.  From the expression of $g_0^{(1)}$ 
in (\ref{zorder}) we see that this term also has $1/y^2$ and $\log(y)$ divergences as
$y\rightarrow 0$ and  needs to be regulated with appropriate counter terms.  
The next order correction in this limit is the 
finite contribution from the constant piece involving $g_1$. 
Thus the behaviour of the Green's function does not satisfy property 2.

To regulate the Green's function so that it satisfies property 2 we   follow
\cite{Romatschke:2009ng} and consider
\begin{equation}
\label{regreen}
\delta G_R(\omega) = G_R(\omega, T) - G_R(\omega, 0)  + 
\frac{N^2}{8\pi^2 L^6}    \frac{r_+^4}{L^2} \frac{6}{5} +P  .
\end{equation}
Let us recall the definition of each of the terms in the above expression. 
\begin{eqnarray}
G_R(\omega, T) = - \frac{N^2}{8\pi^2 L^6} g_R(\omega) -P + G_{\rm{counter}}(\omega, T) .
\end{eqnarray}
Here we have just substituted the expressions for $\hat G_R(\omega) $ and 
$G_{\rm{contact}}$  in the expression for the Green's function given in (\ref{greendef}). 
The $T=0$ Green's function is given by 
\begin{eqnarray}
 G_R(\omega, 0 )  =  \frac{N^2}{8\pi^2 L^6}\frac{r_+^4}{L^2}\lim_{y\rightarrow 0} 
\frac{1}{y^3}  \lambda^4 g_0^{(1)} (y)  + G_{\rm{counter}} ( \omega),   \\ \nonumber
= \frac{N^2}{8\pi^2 L^4 } \omega^4  
\lim_{y\rightarrow 0}\frac{ g_0^{(1)} (y) }{y^3}  + G_{\rm{counter}} ( \omega, 0) .
\end{eqnarray}
Note that there is no contact term in this case since at $T=0$, the pressure vanishes. 
We now point out that we have the following equality
\begin{equation}
G_{\rm{counter}}(\omega, 0) = G_{\rm{counter}} (\omega, T) .
\end{equation}
The reason is clear because the $r^2$ and $\log(r)$  divergences   occur only in the 
term $g_0^{(1)}$  and this term is common both for $T\neq 0$ and $T=0$. 
Thus the counter term to cancel these divergences must be identical. We also can infer that
they are proportional to $\omega^4$. 
Now let us examine the behaviour of $\delta G_R(\omega)$ as $\omega \rightarrow \infty$. 
Substituting all the equations we see that 
\begin{equation}
 \lim_{\omega\rightarrow \infty}\delta G_R(\omega)  = 
\lim_{\lambda\rightarrow\infty} \frac{N^2}{8\pi^2 L^6} \frac{ r_+^4}{4L^2} 
\frac{B(\lambda)}{A(\lambda)} \rightarrow 0.
\end{equation}
Thus $\delta  G_R(\omega)$ satisfies the property 2. 
What has been essentially done by the construction in (\ref{regreen}) is the divergent and constant
pieces of the Green's function as $\omega\rightarrow \infty$ are subtracted out. 
Now that $\delta  G_R(\omega)$  satisfies property 2. we can apply Cauchy's theorem
and obtain the sum rule. 
For later purposes note that 
\begin{eqnarray}
\label{valim}
 {\rm{Im}}\,  \delta G_R(\omega) &=& {\rm{Im}}\,  G_R(\omega, T)  - {\rm{Im}}\,  G_R((\omega, 0),  
\\ \nonumber
&=& \rho(\omega, T) - \rho(\omega, 0)  ,
\end{eqnarray}
since the remaining expressions in (\ref{regreen}) are real.

\subsection{The sum rule}

Now that property 1 as well as property 2 are satisfied by the function $\delta G_R(\omega)$, 
we can evaluate the LHS of the sum rule. This is given by 
\begin{equation}
\label{serf1}
\delta G_R(0)  = G_R(0, T) - G_R(0, 0)  + \frac{N^2}{8\pi^2 L^6}    \frac{r_+^4}{L^2} \frac{6}{5} +P .
\end{equation}
Now we also have
\begin{eqnarray}
\label{zerf}
G_R(0, T) &=& - \frac{N^2}{8\pi^2 L^6} g_R(0) -P,    \\ \nonumber
&=& -P +G_{\rm{counter}}(0, T), \\ \nonumber
G_R(0, 0) &=& 0.
\end{eqnarray}
Here we have used that fact that $g_R(0)=0$ and the fact that 
$G_{\rm counter} $ vanishes at $\omega=0$ since as we have seen earlier that they 
are proportional to $\omega^4$. 
Using the equations (\ref{zerf}) in (\ref{serf1}) we obtain 
\begin{equation}
\delta G_R(0) = \frac{N^2}{8\pi^2 L^6}    \frac{r_+^4}{L^2} \frac{6}{5} = \frac{2}{5}\epsilon.
\end{equation}
Now using (\ref{valim}) and the fact that $\delta G_R(\omega)$ satisfies both
property 1 and property 2 we obtain the shear sum rule derived in \cite{Romatschke:2009ng}
\begin{equation}
\label{ucsum}
\frac{2}{5} \epsilon = \frac{1}{\pi} \int_{-\infty}^\infty  \frac{d\omega}{\omega}
\left( \rho(\omega) - \rho_{T=0} (\omega) \right).  
\end{equation}
Note that the LHS of the sum rule  essentially originates from  the constant term in the 
function $g_R(\omega)$ as $\omega\rightarrow \infty$.

\subsection{Sum rule from OPE}

From the explicit derivation of the sum rule, we see that the LHS side essentially depends
on the high frequency behaviour of the Green's function. In fact it is just proportional to the  constant 
term in the high energy behaviour of the Green's  function. 
Coefficients in the Operator product expansion of the stress tensor contains the information of
the short distance or high energy behaviour of the Green's function.
Thus we should be able to obtain the sum rule from the OPE of the stress tensor
\cite{Romatschke:2009ng}.
This OPE is given by 
\begin{equation}
 T_{\mu\nu}( x) T_{\rho\sigma}(0) \sim C_T \frac{ I_{\mu\nu, \rho\sigma} }{x^8} + 
\hat A_{\mu\nu\rho\sigma\alpha\beta}(x) T_{\alpha\beta}(0) + \cdots ~.
\end{equation}
$\hat A$ contains various Lorentz structures which 
can be found in \cite{Osborn:1993cr}.  The important property 
of them which is seen form conformal invariance is that 
they all scales like $1/x^4$. 
Now we take the Fourier transform of the above OPE and set $q=0$ and  $ \omega\rightarrow \infty$.
By a simple scaling analysis it can be seen that the term proportional to $\frac{C_T}{x^8}$ 
scales like $\omega^4$. This divergence is the same divergence seen in the gravity calculation
due to the terms which are identical to the zero temperature case.  
The term proportional to $\hat A$ gives rise to a constant independent
of $\omega$.  Thus the constant contribution to the Green's function 
as $\omega\rightarrow\infty$ arises due to the 
one point function of the stress tensor in the thermal ensemble. 
Now evaluating 
\begin{equation}
\lim_{\omega\rightarrow  \infty} (  G_R(\omega, T) - G_R(\omega, 0 ) ) =   -P - \frac{2}{5}\epsilon.
\end{equation}
We have used the result obtained in \cite{Romatschke:2009ng} but in the Minkowski space and the convention of \cite{Gulotta:2010cu}. 
From hydrodynamics since the Green's function is defined as the response of the 
changes in the stress tensor on perturbing the metric, the zero frequency value of the 
Green's  function is just given by negative of the pressure. 
\begin{equation}
 G_R(0,  T) = - P, \qquad G_R(\omega, 0) =0. 
\end{equation}
On defining 
\begin{equation}
 \delta G_R(\omega T) =  G_R(\omega, T) - G_R(\omega, 0 ) + P + \frac{2}{5} \epsilon ,
\end{equation}
we see that the LHS side of the sum rule is just the difference between the constant terms in the high
frequency behaviour and the zero frequency behaviour of the Green's function. 
This results in the sum rule given in (\ref{ucsum}).

\section{The shear sum rule for the R-charged D3-brane}

In this section we examine the gravity dual of ${\cal N}=4$ Yang-Mills at
 finite chemical potential and  finite temperature  and re-derive the sum rule. 
Since there are $3$ R-charges corresponding to the Cartan's of $SO(6),$ it is possible to 
turn on $3$ independent chemical potentials. The gravity dual of this system is given by the 
R-charged black hole of Behrndt, Cvetic and Sabra \cite{Behrndt:1998jd}. 
Using the differential equation of the massless minimally coupled scalar in this background,
we obtain the retarded Green's function of the $T_{xy}$ component of the 
stress tensor.  Examining this differential equation and the method discussed 
for the uncharged D3-brane, we will show that the  regulated 
Green's function satisfies both property 1 and property 2 which is necessary for the
derivation of the sum rule. 
We show that the sum rule is corrected by terms which depend on the 3
chemical potentials.  We explain these additional terms in the LHS of the sum rule 
due to the presence of  scalars  in the OPE of the stress tensor. 
The procedure to demonstrate property 1 and property 2 of the retarded 
Green's function parallels that of the uncharged D3-brane, therefore we will be brief and 
highlight only the differences.

\subsection{Green's function from gravity}

We begin with the metric for the R-charged D3-brane with all the three charges turned on
which is given by 
\begin{eqnarray}
\label{chgmet}
 ds^2_5 &=& - { \cal H}^{-2/3}{(\pi T_0 L)^2 \over u}\,f \, dt^2 
+  { \cal H}^{1/3}{(\pi T_0 L)^2 \over u}\, \left( dx^2 + dy^2 + dz^2\right)
+ {\cal H}^{1/3}{L^2 \over 4 f u^2} du^2\, \nonumber  \\ \nonumber
f(u) &=& { \cal H} (u) - u^2  \prod_{i=1}^3 (1+k_i)\,, 
\;\;\;\;\; H_i = 1 + k_i u \,, \;\;\;\;\; 
k_i  \equiv {q_i\over r_H^2}\,, \;\;\;\;\; T_0 =\frac{ r_+}{\pi L^2},
\label{identif} \\ 
u &=& \frac{r_+^2}{r^2}, \quad  {\cal H} = ( 1+ k_1u)( 1+k_2u) (1+k_3u) .
\end{eqnarray}
The scalar fields and the gauge fields in this background are given by 
\begin{equation}
\label{scalval}
 X^i = \frac{ {\cal H}^{1/3} }{ H_i ( u) }, \quad 
A_t^i = \frac{\tilde  k_i \sqrt{2} u}{ L H_i(u) }, \quad
\tilde k_i = \sqrt{q_i}{L }\prod_{i=1}^3 ( 1 + k_i)^{1/2} .
\end{equation}
The above metric is a  solution of the equations of motion of the STU-model given by
action  
\begin{eqnarray}
\label{cgaction}
 S &=& \frac{N^2}{8\pi^2 L^3} \int d^5 x\sqrt{-g} {\cal  L}, \\ \nonumber
{\cal L} &=& R + \frac{2}{L^2} {\cal V} - \frac{1}{2} G_{ij} F^{i}_{\mu\nu} F^{\mu\nu\, j}-
G_{ij} \partial_\mu X^i \partial^\mu X^j + \frac{1}{24 \sqrt{-g} }
\epsilon^{\mu\nu\rho\sigma\lambda} \epsilon_{ijk} F_{\mu\nu}^i F^{\rho\sigma j} A^k_\lambda ,
\end{eqnarray}
where $F^{i}_{\mu\nu}, i =1, 2, 3$ are the field-strengths for the 
three Abelian gauge fields. The three scalar fields $X^i$'s are subject to the 
constraint $X^1X^2X^3 =1$. The metric on the scalar manifold is given by 
\begin{equation}
 G_{ij} = \frac{1}{2} {\rm diag} \left\{ (X^{1})^{-2},  (X^{2})^{-2},  (X^{3})^{-2} \right\}.
\end{equation}
The scalar potential is given by 
\begin{equation}
 {\cal V} = 2 \left( \frac{1}{X^1} +   \frac{1}{X^2} + \frac{1}{X^3} \right) .
\end{equation}
It is useful to recall the following thermodynamic data of this black hole \cite{Son:2006em}. 
The Hawking temperature $T_H$, entropy density $s$ , energy density $\epsilon$, pressure $P$, 
charge densities $\rho_i$  and the conjugate chemical potentials $\mu_i$  are given by 
\begin{eqnarray}
\label{cthermv}
 T_H = \frac{ 2 +k_1 + k_2 + k_3 - k_1k_2k_3}
{ 2 \sqrt{( 1+ k_1) ( 1+ k_2) ( 1+ k_3) } } T_0, &\qquad& 
s =  \frac{\pi^2 N^2 T_0^3}{2} \prod_{i=1}^3 ( 1 + k_i)^{1/2}, 
\\ \nonumber
\epsilon = \frac{ 3\pi^2 N^2 T_0^4}{8} \prod_{i=1}^3 ( 1 + k_i) ,  &\qquad& 
P = \frac{\pi^2 N^2 T_0^4}{ 8}  \prod_{i=1}^3 ( 1 + k_i), \\ \nonumber
\rho_i = \frac{\pi N^2T_0^3}{8} \sqrt{2k_i}  \prod_{i=1}^3 ( 1 + k_i),  &\qquad& 
\mu_i = \frac{\pi T_0 \sqrt{2k_i}}{( 1+ k_i)}  \prod_{i=1}^3 ( 1 + k_i). 
\end{eqnarray}
The thermodynamical stability condition of this black hole is given by 
\begin{equation}
 2 -k_1-k_2 - k_3 + k_1 k_2k_3>0.
\end{equation}

As before, to study the retarded correlator $G_R(\omega, T)$ we must examine the equation of 
motion of a massless minimal coupled scalar in this background. 
We consider the  perturbation 
\begin{equation}
 \delta g_{xy} = \phi(r) {\cal H}^{1/3} \frac{r^2}{L^2} e^{-i\omega t},
\end{equation}
whose equations of motion  in radial co-ordinate $r$  is given by   \cite{Son:2006em}
\begin{equation}
\label{mceq1}
\partial_r^2 \phi +\left( \frac{F'}{F} + \frac{3}{r} \right) \partial_r \phi + 
\frac{{\cal H} \omega^2}{F^2} \phi =0,
\end{equation}
where again $F$ is given by 
\begin{equation}
 F = \frac{r^2}{L^2} f 
\end{equation}
and $f$ is defined in (\ref{chgmet}). 
This equation is similar to the uncharged case except for the presence of ${\cal H}$ as coefficient
of $\omega^2$ and the fact that $f$ depends on the constants $k_i$. 
To obtain the retarded Green's function, we must impose in-going boundary conditions at the horizon
$r_+$ and obtain $\phi$ at the boundary $r\rightarrow \infty$. Then the Green's function 
is given by the expression in (\ref{greendef}). 
$G_{\rm{contact}}$ is the contribution of the contact terms obtained from the on shell
effective action. For the charged case, 
this can be evaluated by using following action 
\begin{equation}
S =  \frac{N^2}{8\pi^2 L^3} \int d^5 x\sqrt{-g}  {\cal L} + \frac{N^2}{4\pi^2 L^3 }\int_{\partial M_5}
d^4 x \sqrt{-h} K + \frac{N^2}{4\pi^2 L^3} \int_{\partial M_5} \sqrt{-h} W.
\end{equation}
Here the second term is the Gibbons-Hawking boundary term while the third term is the 
term required to make the action finite in the limit $r\rightarrow\infty$ \cite{Batrachenko:2004fd}. It is given by
\begin{equation}
\label{bcounterc}
W = - \frac{{\cal H}^{1/3}}{L }\sum_{i =1}^3 H_i^{-1}.
\end{equation}
Using this, it can be shown that the  contact terms are again obtained from the 
on shell action given in (\ref{os}) \cite{Son:2006em}
\footnote{See equation 4.24 of \cite{Son:2006em} }.  
Therefore we obtain 
\begin{equation}
G_{\rm{contact}} = -P.
\end{equation}
There is an alternative indirect method to infer that the contact term is given by $-P$. 
Contact terms do not depend on frequency, therefore one can obtain them by taking the zero frequency
limit. It will be shown that $\hat G(\omega, T)\rightarrow 0$ in the $\omega\rightarrow 0$ limit, further more
the $G_{\rm{counter}}$ as in the case of the uncharged case behave as $\omega^4$. Thus in the 
zero frequency limit the only contribution to the Green's function is from $G_{\rm{contact}}$. 
Now from hydrodynamics, the Greens' function is the response of the stress tensor $T_{xy}$ to 
the metric fluctuation $\delta g_{xy}$. At zero frequency, this is just $-P$. 
Thus we have $G_{\rm{contact}} = -P$.

Before we begin our analysis,  we will need the behaviour of the solutions near the horizon
and the boundary. 
The behaviour of the solutions near the horizon is given by 
\begin{equation}
\label{nrhor}
 \phi(r) \sim ( r - r_+) ^ {\pm i  \alpha \omega} , 
\end{equation}
where $\alpha$ can be written as
\begin{eqnarray}
 \alpha &=& \frac{1}{F_h} \sqrt{( 1+k_1)(1+k_2)(1+k_3)} \\ \nonumber
  F &=& ( r-r_h) F_h + \cdots , \qquad 
F_h =  \frac{2 r_+}{L^2} ( 2 + k_1 + k_2 + k_3 - k_1k_2k_3). 
\end{eqnarray}
As $r\rightarrow\infty$, the metric asymptotes to $AdS_5$, therefore the behaviour of 
$\phi(r)$ is the same as in the uncharged case and is given by (\ref{bexp}). 
Finally the equations for $\phi(r)$ can be obtained by variation of the following action 
\begin{equation}
\label{varact2}
S_\phi  = \int _{r_h}^\infty dr Fr^3 \left( 
|\phi'(r)|^2 - \frac{\omega^2{\cal H}}{F^2} |\phi(r)|^2 \right) . 
\end{equation}
The only change in this action compared to the uncharged case 
in (\ref{varact}) is the presence of ${\cal H}$ in the term proportional to the frequency.

\subsection{Green's function in the $\omega$-plane}

In this subsection we will discuss the analytic properties of the function 
\begin{equation}
\label{defgr1}
 g_R(\omega) = \lim_{r\rightarrow\infty} \frac{Fr^3\phi'}{\phi}, 
\end{equation}
in the complex $\omega$ plane.  The discussion closely follows that of the 
uncharged case.  The difference arises mainly in the behaviour of the $\omega\rightarrow\infty$
limit which we will highlight.

\vspace{.5cm}
\noindent
{\bf  No poles for ${\rm Im} \, \omega >0$ }
\vspace{.5cm}

Here the discussion is identical to that of the uncharged case. 
Poles or divergences  in $g_R(\omega)$ correspond to quasi-normal modes of the equation 
(\ref{mceq}).  Quasi-normal modes are solutions to the equation 
(\ref{mceq}) with the following boundary conditions
\begin{eqnarray}
 \phi(r) \sim ( r-r_+)e^{-i\alpha \omega}, \qquad r\rightarrow r_+, \\ \nonumber
\phi(r) \sim r^{-4}, \qquad r\rightarrow \infty.
\end{eqnarray}
 One can follow the same arguments as in the case of the uncharged situation to show
that such modes do not exist for ${\rm Im} \, \omega >0$.
We need to employ the action given in (\ref{varact2}) for this purpose. 
The fact that    ${\cal H}> 0$ in the domain $r_+<  r< \infty$ also  ensures that the 
action $S_{\phi}$ evaluated on the quasi-normal mode is positive definite. 
This enables us to repeat the same arguments as in the case of zero chemical potential. 
 
\vspace{.5cm}
\noindent
{\bf  No poles of $\omega$ real and $\omega\neq 0$}
\vspace{.5cm}

Here again the same reasoning developed for the uncharged case earlier
using the Wronskian, ensures that  quasi-normal modes do not exist on the real line $\omega\neq 0$.
This implies that there are no poles or divergences  in $g_R(\omega)$ in this domain. 

\vspace{.5cm}
\noindent
{\bf No poles for $\omega =0$}
\vspace{.5cm}

The behaviour of the function $g_R$ as $\omega\rightarrow 0$ can be determined from the
differential equation (\ref{mceq1}) as follows. We define
\begin{equation}
\tilde g(r) = \frac{\phi'}{\omega\phi(r)}.
\end{equation}
Using the equation (\ref{mceq1}) we find the $\tilde g(r)$ satisfies the equation
\begin{equation}
\label{raeq}
\tilde g' + \omega \tilde g^2 + \left( \frac{F'}{F} + \frac{3}{r} \right) \tilde g(r) +
 \frac{\omega {\cal H}}{F^2}=0.
 \end{equation}
 The ingoing boundary condition implies that  $\tilde g$ satisfies
 \begin{equation}
 \tilde g \rightarrow \frac{-i \sqrt{( 1+k_1)(1+k_2)(1+k_3)} }{F_h( r-r_+)}, \qquad r\rightarrow r_+.
 \end{equation}
 The solution of (\ref{raeq}) in the $\omega\rightarrow 0$ limit which satisfies the above boundary condition
 is given by
 \begin{equation}
 \tilde g = -i \frac{r_+^3 \sqrt{( 1+k_1)(1+k_2)(1+k_3)}}{r^3 F}, \qquad \omega\rightarrow0.
 \end{equation}
 This implies that 
 \begin{equation}
 \lim_{\omega\rightarrow 0} g_R(\omega) = \omega \lim_{r\rightarrow \infty} r^3 F\tilde g = 
 -i\omega r_+^3  \sqrt{( 1+k_1)(1+k_2)(1+k_3)}.
 \end{equation}
 Thus $g_R(\omega)$ is proportional to $\omega$ as $\omega\rightarrow 0$. 
Examining the equation (\ref{raeq}), it can be seen that it admits a power series expansion in 
$\omega$ around the origin.

\vspace{.5cm}
\noindent
{\bf No zeros for ${\rm Im}\, \omega>0, \omega \neq 0$}
\vspace{.5cm}

As argued for the uncharged case, zeros or points 
at  which the  Green's function vanish correspond to the boundary condition
\begin{eqnarray}
 \phi(r) &\sim& (r-r_+)^{-i\alpha \omega}, \qquad r\rightarrow r_+, \\ \nonumber
r^3 F\phi'(r) &\rightarrow 0&, \qquad r\rightarrow \infty.
\end{eqnarray}
Using the same reasoning discussed for the uncharged situation,  it is easy to see that there 
are no  modes which have the above boundary condition for ${\rm Im}\, \omega>0, \omega \neq 0$.
This implies that there are no zeros  or points at which the Green's function vanishes in this domain. 

\vspace{.5cm}
\noindent
{\bf No branch cuts for ${\rm Im }\, \omega \geq  0$}
\vspace{.5cm}

From the discussion above, we have seen that the Green's function does not have any singularities, 
nor does it vanish any where in the upper half plane except at the origin. 
At the origin it admits an  analytic power series expansion. 
As we have argued for the uncharged $D3$-brane case the general theorem in (\cite{Arnold})
ensures that the local Forbenius expansion of the solution is smooth with respect to the
parameter $\omega$   
since the differential equation and the boundary condition are both smooth with 
respect to this parameter $\omega$. 
Now from (\ref{defgr1}) 
 it is easy to see that the $n^{\rm th}$ order derivatives of the Green's function with respect to the 
parameter $\omega$ exist. This is because  the singularities for the $n^{\rm th}$ 
order derivative can arise  only from the zeros of $\phi$ which are its  quasi-normal modes. 
We have already  shown that there are no quasi-normal modes 
 in the upper half-plane, hence we  conclude that arbitrary derivatives of 
 the  Green's function  with respect to $\omega$ exists in the upper half plane.
 Thus the Green's function does not 
have any branch cuts in the upper half-plane 

 Therefore  the expansion at $\omega =0$ can be extended to the entire upper half plane 
 and the function is analytic in the upper half plane. 
Thus the Green's function satisfies the property 1 for the derivation of the sum rule.

\vspace{.5cm}
\noindent
{\bf  Behaviour as $\omega \rightarrow \infty$}
\vspace{.5cm}

To obtain the asymptotic behaviour of the Green's function it is again convenient to 
work in the scaled variables
\begin{equation}
 z = \frac{r_+}{r}, \qquad i\lambda = \frac{L^2}{r_+}\omega.
\end{equation}
We then expand the equation (\ref{mceq1}) in powers of $\frac{1}{\lambda}$. For this it is 
convenient to express the background metric given in (\ref{chgmet})  in 
 Fefferman-Graham coordinates. 
The metric for the charged D3-brane  
in Fefferman-Graham coordinates is given by 
\begin{equation}
\label{fgexpm}
 ds^2 = \frac{L^2}{ z^2} \left( ( -1 + \sum_{i=2}^\infty a_i (z^2)^{i} ) dt^2 
+ ( 1 + \sum_{i=2}^\infty b_i ( z^2)^{i} ) ( d\vec x^2 ) + dz^2 \right) .
\end{equation}
The details of the transformation to these co-ordinates are given in the appendix. 
Note that in this expansion the term $a_1, b_1$ is missing.  Here the boundary is at 
$z=0$.  It will be seen that the coefficients $a_2, b_2$ are crucial to determine the 
asymptotic behaviour of the Green's function.  These  are evaluated in (\ref{gnfg}). 
 The equation for the minimally coupled scalar in this co-ordinates is given by 
\begin{equation}
 \phi'' -\frac{1}{z} \left( 3 + \sum_{i=2}^\infty \tilde a_i (z^2)^i \right) \phi'
 - \omega^2 \left( 1 + \sum_{i =2}^\infty \tilde b_i (z^2)^i \right) \phi =0,
\end{equation} where
 $\tilde a_i, \tilde b_i$ are related to the coefficients $a_i, b_i.$
Again the leading coefficients are important for our later use which are related by 
\begin{equation}
\label{abrel}
 \tilde a_2 = 2 ( a_2 -3b_2) , \qquad \tilde b_2 = a_2.
\end{equation}
We now rescale the co-ordinate $z$ by defining 
\begin{equation}
 y = \lambda z. 
\end{equation}
This leads to the equation for the minimally coupled scalar as a $\frac{1}{\lambda}$ expansion. 
\begin{equation}
\label{scaleq}
 \phi''(y) - \frac{1}{y} \left( 3 + \sum_{i=2}^\infty \tilde 
a_i (\frac{y^2}{\lambda^2}) ^i \right) \phi'(y) - 
( 1 + \sum_{i =2}^\infty \tilde b_i ( \frac{y^2}{\lambda^2})^i ) \phi( y) =0.
\end{equation}
Now this expansion is similar to that
obtained for the un-charged situation given in (\ref{lamexp}).   We can  therefore 
use the same method to solve it 
 order by order in $1/\lambda$. We define
\begin{equation}
 g(y)  = \frac{\phi'(y)}{\phi(y) }.
\end{equation}
From  the equation given in (\ref{scaleq}) we see that the $g$ admits an expansion 
of the form
\begin{equation}
 g = g_0  + \frac{1}{\lambda^4} g_1 + \frac{1}{\lambda^6} g_2 + \cdots.
\end{equation}
The equations satisfied by these can be obtain from  (\ref{scaleq})
and the first few orders are given by 
\begin{eqnarray}
\label{perteq}
& &  g_0'  + g_0^2  - \frac{3}{y} -1 =0, \\ 
& & g_1' + \left( 2 g_0 - \frac{3}{y} \right) g_1 - \tilde a_2 y^3 g_0 - \tilde b_2 y^4 =0, \\
& & g_2' + \left( 2 g_0 - \frac{3}{y} \right) g_2 - \tilde a_3 y^5 g_0 - \tilde b_3 y^6 = 0 .
\end{eqnarray}
We will now follow the same procedure as done for the uncharged case to 
obtain two independent solutions to (\ref{scaleq}). 
The zeroth order solution is same as the uncharged case since in the $\lambda\rightarrow \infty$
limit, the equation (\ref{scaleq}) reduces to the $\lambda\rightarrow\infty$ limit
of (\ref{lamexp}). This is expected since asymptotically the metric (\ref{chgmet}) 
 reduces to $AdS_5$.  Therefore we have
\begin{eqnarray}
 g_0^{(1)} = -\f{K_1(y)}{K_2(y)}= \f{d}{dy}\left(  \log(y^{2}K_{2}(y) ) \right) , \;\;
g_0^{(2)} = \f{I_1(y)}{I_2(y)}= \f{d}{dy}\left( \log(y^{2}I_{2}(y))\right). 
\end{eqnarray}
Let us first find the $\frac{1}{\lambda}$ expansion around the solution $g_0^{(1)}$. 
The first order correction is given by 
\begin{equation}
 g_1^{(1)}(y) = \frac{1}{y K_2^2(y)} \int_0^y dy \left( \tilde a_2 y^3 g_0 + \tilde b_2 y^4 \right)
y K_2^2(y)  + \frac{c_1}{y K_2^2(y) }.
\end{equation}
We set $c_1=0,$ so as  not to change the asymptotics of the solution at $y\rightarrow\infty$.  
On performing the integral we obtain
\begin{eqnarray}
\label{cg11}
 g_1^{(1)}( y) &=&  \frac{\tilde a_2 }{2} y^3 + \frac{\tilde b_2}{10} y^5( 1 - \frac{ K_3^2}{K_2^2} ),
 \\  \nonumber
&=& \left( \frac{\tilde a_2}{2} - \frac{8}{5} \tilde b_2 \right) y^3 
- \frac{3}{10}  \tilde b_2 y^5 + O( y^7, y^7 \log (y) )  .
\end{eqnarray}
Similarly, the solution to $g_2$ is given by 
\begin{equation}
 g_2^{(1)}(y) = \frac{1}{y K_2^2(y)} \int_0^y \left( \tilde a_3 y^5 g_0 + \tilde b_3 y^6 \right)
y K_2^2(y) .
\end{equation}
From the expansions of the functions involved, one sees that
\begin{equation}
 g_2^{(1)}(y) =  y^7 + O(y^9, y^9 \log y). 
\end{equation}
Similarly from the structure of the functions in the $\frac{1}{\lambda}$ expansion, it is easy to see that 
\begin{equation}
 g_n(y) \sim y^m, \qquad  m \geq 7, \quad \hbox{for} \;\;n \geq 2, \quad y \rightarrow 0.
\end{equation}
Using this procedure it is possible  to construct all the $g_n^{(1)} $'s. Form this 
we find that the first solution of the differential equation is given by 
\begin{eqnarray}
\label{chphi1}
 \phi^{(1)} (y) &=& \exp( \int_0^y dy \left( g_0 + \frac{g_1}{\lambda^4}  + \frac{g_2}{\lambda^6}
+ \cdots \right) , \\ \nonumber
&=& y^2 K_2(y) \left( 1 + \frac{1}{\lambda^4} \int_0^y dy g_1(y)  + \frac{1}{\lambda^6} 
\int_0^y dy g_2 (y) + \cdots \right).
\end{eqnarray}
Note that 
\begin{equation}
\phi^{(1)} \sim {\rm constant}, \quad y \rightarrow 0.
\end{equation}
  From (\ref{cg11})  
we  see that $\frac{1}{\lambda^4}$ term in (\ref{chphi1})  goes as $y^4$  near the origin.
The higher order terms are further suppressed as $y\rightarrow 0$.  
A similar construction can be done starting with the seed $g_0^{(2)}$. 
The first order term about this seed is 
\begin{eqnarray}
\label{cg21}
 g_1^{(2)}(y) &=& \frac{1}{y I_2^2(y) } \int_0^y dy \left(  \tilde a_2 y^3 g_0 + \tilde b_2 y^4 \right),  \\ \nonumber
&=& \frac{\tilde a_2 }{2} y^3 + \frac{\tilde b_2}{10} y^5( 1 - \frac{ I_3^2}{I_2^2} ).
\end{eqnarray}
Similarly  $g_2^{(2)}(y)$ can be written as
\begin{equation}
 g_2^{(2)}(y) = \frac{1}{y I_2^2(y)} \int_0^y \left( \tilde a_3 y^5 g_0 + \tilde b_3 y^6 \right)
y I_2^2(y) .
\end{equation}
 Examining  this solution, we see that  the behaviour of this function as $y\rightarrow 0$ is given by 
\begin{equation}
 g_2^{(2)}(y) \sim y^5 +O(y^7), \qquad y\rightarrow 0.
\end{equation}
From the general form of the equations for $g_n,$ it can be seen that 
\begin{equation}
\label{cg2n}
 g_n^{(2)} (y) \sim y^m , \qquad m\geq 5 \qquad \hbox{for}\quad n\geq 2, \quad y\rightarrow 0.
\end{equation}
Thus one obtains the solution 
\begin{equation}
 \phi^{(2)} (y) = y^2 I_2(y) \left( 1 + \frac{1}{\lambda^4} \int _0^y g_1(y) + \frac{1}{\lambda^6}
\int_0^y g_2(y)  + \cdots \right) .
\end{equation}
It is important to observe that  
\begin{equation}
\phi^{(2)}(y) \sim y^4, \qquad y\rightarrow 0,
\end{equation}which is same as in the case of the uncharged case.
Again from (\ref{cg21}) and (\ref{cg2n}), the terms in the $\frac{1}{\lambda}$ expansion are 
suppressed as $y\rightarrow 0$. 

Once the two independent solutions have been constructed in this manner, we can write the
solutions which obey the boundary conditions as 
\begin{equation}
 \phi(y) = A(\lambda)  \phi^{(1)} (y) + B(\lambda)  \phi^{(2)}.
\end{equation}
The coefficients $A$, $B$ depend on $\lambda$ and can be obtained by matching $\phi(y)$ and 
its derivative  with the near solution.
The near solution is obtained in terms of a  
Forbenius power series expansion at the horizon. The solution is of the form 
\begin{equation}
 \phi = ( 1-\frac{y}{\lambda} ) ^{\frac{\alpha \lambda r_+}{L^2}} 
\left( \sum_{j=0}^\infty a_j ( 1- \frac{y}{\lambda})^j \right).
\end{equation}
Now though we do not need the detail dependence of the coefficient $B$ on $\lambda$, it is 
clear that  using the same argument as in the uncharged situation  we find
\begin{equation}
\label{bazero}
 B(\lambda)  \rightarrow 0, \qquad 
 A(\lambda) \rightarrow \frac{1}{2}, \quad \hbox{as} \quad \lambda\rightarrow \infty .
\end{equation}
The reason is that we know that at $\lambda\rightarrow \infty$, the equation reduces to that 
of the $T=0$  limit.  In this 
situation   the solution which is finite at $y\rightarrow \infty$ is 
$\frac{1}{2} y^2 K_2(y)$. 

With this input  and using the same steps followed in the uncharged situation we find
the Green's function to be 
\begin{equation}
 g_R(\lambda)   = - \frac{r_+^4}{L^2} \left(  \lim_{y\rightarrow 0} \frac{\lambda^4}{y^3} 
g_0^{(1)}(y)   - \frac{3}{5} ( a_2 + 5 b_2)  + \frac{1}{4}  \frac{ B(\lambda)}{ A(\lambda)} \right).
\end{equation}
In this we have substituted for $\tilde a_2, \tilde b_2$ from (\ref{abrel}). 
Again, the above result is valid for all values of $\lambda$. Taking the $\lambda\rightarrow \infty$
limit we are left with 
\begin{equation}
 \lim_{\lambda\rightarrow \infty}  g_R(\lambda)  = - \frac{r_+^4}{L^2}  
\left(  \lim_{y\rightarrow 0}\frac{\lambda^4 g_0^{(1)}(y) }{y^3}  - \frac{3}{5} ( a_2 + 5 b_2)  \right),
\end{equation}
where we have used (\ref{bazero}).  As we have seen 
earlier the $\lambda^4$ contribution is identical to the zero temperature  case.  There is 
also a constant term due to the first order correction $g^{(1)}$. 
Thus the Green's function does not satisfy property 2 and needs to be regulated. 

As before, to regulate the Green's function we consider
\begin{equation}
 \delta G_R(\omega) = G_R(\omega, T) - G_R(\omega, 0)  + 
\frac{N^2}{8\pi ^2 L^6} \frac{r_+^4}{L^2}\left(  \frac{3}{5} ( a_2 + 5 b_2) \right)  +P.
\end{equation}
Now using all our previous results   and following the same 
steps as in the case of the uncharged situation we have 
\begin{equation}
  \delta G_R(\omega) \rightarrow 0, \qquad \omega\rightarrow \infty.
\end{equation}
We have essentially subtracted the divergent and the constant term in $G_R(\omega)$ so that 
property 2 is true on $\delta G_R(\omega)$ which can then be used to obtain 
the sum rule.  Note that for the charged D3-brane case  the equation 
(\ref{valim}) still holds since all the terms subtracted to construct $\delta G_R(\omega)$ 
are still real.

\subsection{The sum rule}

We can now use Cauchy's theorem on the function $\delta G_R(\omega)$ and derive the sum rule. 
The sum rule for the charged D3-brane then is given by 
\begin{equation}
 \delta G_R( 0) = \int_{-\infty}^\infty \frac{d\omega}{\omega} 
\left( \rho(\omega) - \rho_{T=0}( \omega) \right),
\end{equation}
where 
\begin{equation}
\label{sumrulec}
\delta G_R(0)  =  \frac{N^2}{8\pi ^2 L^6} \frac{r_+^4}{L^2}\left( \frac{3}{5} (  a_2 + 5 b_2) \right).
\end{equation}
We can now substitute the values of $a_2, b_2$ from (\ref{gnfg}). We obtain
\begin{eqnarray} 
\label{sumcrule}
 \delta G_R(0)  &=&  \frac{N^2}{8\pi ^2 L^6} \frac{r_+^4}{L^2}\frac{3}{5}\left(
2 ( 1+ k_1) ( 1+ k_2) (1+k_3)\right.  \\ \nonumber
& & \left.  -\frac{1}{9} \left\{ ( k_1 - k_2)^2 + ( k_1- k_3)^2 + ( k_2- k_3)^2 \right\} \right)
\\ \nonumber
&=& \frac{2}{5}\epsilon - \frac{N^2\pi^2 T_0^4}{120} \left\{ ( k_1 - k_2)^2 + ( k_1- k_3)^2 + ( k_2- k_3)^2 \right\}.
\end{eqnarray}
Here we have rewritten the first term in terms of the energy density using its expression
given in (\ref{cthermv}). 
Note that there is a modification of the sum rule which involves the chemical potential. 
In the next section we will show that this modification is due to the 
expectation values of the scalars in the charged D3-brane background. 
As a simple check of this hypothesis, note that for D3-brane with equal R-charge in 
  all the 3 Cartan directions $k_1=k_2=k_3$, the additional term vanishes.

\subsection{Sum rule from OPE}

We have seen that the term on the LHS of the sum rule is essential due to the high frequency behaviour
of the retarded Green's function.  In the uncharged situation, 
this high frequency behaviour can be captured by the OPE of the $T_{xy}$ component
of the stress tensor.  Therefore if the sum rule has to be modified there must be 
additional terms in the OPE which contribute at finite chemical potential. 
Let us now re-examine the OPE of the stress tensor which is given by 
\begin{equation}
\label{opeop}
 T_{\mu\nu}( x) T_{\rho\sigma}(0) \sim C_T \frac{ I_{\mu\nu, \rho\sigma} }{x^8} + 
\hat A_{\mu\nu\rho\sigma\alpha\beta}(x) T_{\alpha\beta}(0) + 
B_{\mu\nu\rho\sigma}^a (x) {\cal O}_a (0). 
\end{equation}
Here we have included  additional terms in the OPE which arise if there
are scalars denoted by ${\cal O}_a$ in the theory and the 
three point function $\langle T_{\mu\nu} T_{\rho\sigma} {\cal O}_a \rangle $ is non-zero. 
From a simple scaling analysis it can be  shown that operators of  dimension  lower than 4 in 
the OPE contribute to divergences as $\omega\rightarrow \infty$.  
Operators of dimension greater than $4$ are irrelevant and 
operators of dimension $4$ contribute to the finite terms responsible for the sum rule.  
Consider an operator ${\cal O}_a$  of dimension $\Delta$ in OPE, the structure of this term will be of
the form
\begin{equation}
 B_{\mu\nu\rho\sigma}^a(x) {\cal O}_a(0),
\end{equation}where
$a$ refers both to the operator as well as tensor indices if any. 
By conformal invariance we have the following scaling property
\begin{equation}
B_{\mu\nu\rho\sigma}^a( \Lambda x) = \Lambda^{\Delta-8} B_{\mu\nu\rho\sigma}^a (x).
\end{equation}
Now, on taking the Fourier transform of this term   
we obtain
\begin{equation}
\label{cisc}
 \int d^4x  e^{-i\omega t} B_{\mu\nu\rho\sigma }^a( x) {\cal O}_a(0) = 
\omega^{4-\Delta}  \int d^4 \tilde x e^{-i \tilde t} B_{\mu\nu\rho\sigma  }^a ( \tilde x)  {\cal O}_a(0) .
\end{equation}
Here we have rescaled all coordinates by replacing $x= \frac{\tilde x}{\omega}$. 
From (\ref{cisc}) we see that the terms diverge for $\Delta <4$  and are 
irrelevant for $\Delta >4$. The finite terms in the OPE are therefore coming
from operators of dimension $\Delta =4$. 

We have seen from the discussion of the uncharged D3-branes, the term proportional
to the stress tensor  contributes to term $\frac{2}{5}\epsilon$ in the sum rule.  Thus the remaining
terms in (\ref{sumcrule}) must be due to the expectation values of certain dimension $4$  operators. 
From the structure of the extra  terms in  (\ref{sumcrule}), we see that they vanish
when all the charges are equal.   At this special point one can see from the 
solution (\ref{scalval}), the scalars become trivial while the gauge field still remains non-zero. 
Thus it  must be the case  that operators dual to the scalars are  responsible for the 
additional terms in  the sum rule. 
To find the conformal dimensions of the operators dual to these scalars we evaluate their masses. 
Due to the constraint $X_1X_2X_3 =1$, there are 
 two independent scalars. 
We parametrize  $X_i$'s in terms of these independent scalars by   
\begin{eqnarray}
\label{xphi1c}
& & X_1 = \exp\left(  -\frac{1}{\sqrt{6}} \phi_1- \frac{1}{\sqrt{2}}  \phi_2 \right) ,\qquad 
X_2 = \exp\left(   -\frac{1}{\sqrt{6}}\phi_1+ \frac{1}{\sqrt{2}}  \phi_2    \right), \\ \nonumber 
& &  X_3 = \exp\left( \frac{2}{\sqrt{6}} \phi_1 \right) .
\end{eqnarray}
Substituting these redefinitions in the action (\ref{cgaction}) and expanding the scalar potential 
\begin{equation}
 V =\frac{4}{L^2}\sum_{i=1}^{3}\frac{1}{X_i}, 
\end{equation}
to quadratic order in $\phi_i$ it can be shown that the masses of both $\phi_1, \phi_2$ is given by 
\begin{equation}
 m^2 L^2 =-4.
\end{equation}
From the   mass-dimension relation 
\begin{equation}
 \Delta( \Delta -4) = m^2L^2, 
\end{equation}
we see that the fields $ \phi_i$ corresponds to operators of dimension $2$ in the field theory and 
saturate the Breitenlohner-Freedman bound.  
In the  field theory  these operators correspond to  two linear combinations  of the 
following chiral primaries
\begin{equation} 
{\rm Tr} ( X\bar X), \qquad  { \rm Tr} ( Y\bar Y), \qquad  {\rm Tr}(Z\bar Z) , 
\end{equation}
where $X, Y, Z$ are 3 complex fields constructed out of the $6$ scalars in ${\cal N}=4$ Yang-Mills. 
These are the three scalars which are un-charged under the Cartans of $SO(6)$. 
Note that the combination $ {\rm Tr} ( X\bar X) +   { \rm Tr} ( Y\bar Y) +   {\rm Tr}(Z\bar Z) $
is the Konishi scalar and therefore not a  chiral primary. 
Let us denote the two chiral primaries dual to the field $\phi_1, \phi_2$ as 
${ O}_1$ and ${O}_2$ respectively. 

From the gravity background it is easy to read out the expectation values of the 
operators ${O}_1$ and ${ O}_2$  at finite chemical potential. 
 Since these fields saturate the Breitenlohner-Freedman bound their behaviour at the boundary
is given by 
\begin{equation}
 \phi_i(r) \sim L \left( \alpha_i \frac{\log r}{r^2 }  + \beta_i \frac{1}{r^2}  + \cdots \right).
\end{equation}
Here we have introduced factors of $L$ so as to ensure that $\alpha_i, \beta_i$ have the 
appropriate dimensions. 
The boundary term for the scalars given in (\ref{bcounterc}) 
 ensures that $\alpha_i$ is set to zero at the boundary.  Then the expectation value of 
the operator dual to $\phi_i$
is obtained  by substituting the redefinitions (\ref{xphi1c}) and in 
the background value of $X_i$ and reading out the value of $\frac{N^2}{8 \pi^2 L^3}\beta_i$ 
\cite{Marolf:2006nd}. See \cite{Witten:2001ua, Berkooz:2002ug} for earlier work.  
They are given by 
\begin{eqnarray}
\label{expect}
  \langle O_1\rangle &=&  \frac{N^2}{8 \pi^2 L^3}\beta_1 = 
\frac{N^2}{8\pi^2 } \frac{r_+^2}{L^4}  \frac{k_1+k_2-2k_3}{\sqrt{6}}, \\ \nonumber
\langle O_2 \rangle &=& \frac{N^2}{8 \pi^2 L^3}\beta_2 = 
\frac{N^2} {8\pi^2 }\frac{r_+^2}{L^4} \frac{k_1-k_2}{\sqrt{2}}.
\end{eqnarray}
Note that these expectation values have mass dimension 2 since the operators $O_1, O_2$ have
mass dimension 2. They are also constant in space time. 

A natural candidate for the operators ${\cal O}^i$ of dimension 4 which occur in the OPE 
and which can contribute to the sum rule 
are the squares of the operators \footnote{The operators 
$O_i$  do not occur directly in the OPE  since the 3 point function
$\langle T_{\mu\nu} T_{\rho\sigma} O_i \rangle$ vanishes in gravity.} 
\begin{equation}
{\cal O}_1 = 
\frac{1}{N}O_1^2, \qquad {\cal O}_2 = 
\frac{1}{N}O_2^2, \qquad {\cal O}_3 =  \frac{1}{N}O_1O_2. 
\end{equation}
We have defined these operators with the factor of $\frac{1}{N}$ so that the two point 
function of all these operators are 
normalized as $N^2$. 

We first show that the OPE of the stress tensor can give rise to these operators. 
The operators $O_1$ and $O_2$ are  chiral primaries, and they will contain 
say the operator ${\rm Tr} ( Z\bar Z)$. Thus the operator ${\cal O}_1$ and  ${\cal O}_2$
will contain the operator 
\begin{equation}
{\cal P}=  \frac{1}{N}(  {\rm Tr} ( Z\bar Z))^2   
\end{equation} 
Therefore we evaluate the following   three point function 
in the free field limit
\begin{eqnarray}
\label{freefield}
 \langle T_{xy}(x^1)T_{xy}(x^2) {\cal P}(x^3)  \rangle 
=  \frac{1}{N^3} \langle {\rm Tr}( \partial_{{x}^{1}}Z \partial_{{y}^{1}}\bar{Z} )
{\rm Tr}( \partial_{{x}^{2}}Z\partial_{{x}^{2}}\bar{Z}^{2} )
{\rm Tr} (Z\bar{Z} )  {\rm Tr}(  Z \bar{Z})  \rangle 
\end{eqnarray}
To obtain this equality we have written out only the terms which contribute to the 
correlator in the free field limit. We have also ignored numerical proportionality constants. 
Note that the two point function of the stress tensor is normalized to unity. 
By applying Wick's contractions  we obtain
\begin{eqnarray}
\langle T_{xy}( x^1) T_{xy}(x^2)  {\cal P}(x^3) \rangle  = 
N \frac { (x^{1}-x^{3})(y^{1}-y^{3})(x^{2}-x^{3})(y^{2}-y^{3})}
{|x^{1}-x^{3}|^{8}|x^{2}-x^{3}|^{8}}
\end{eqnarray}
Therefore from the free field calculation it is evident that
\begin{equation}
\langle T_{xy}T_{xy} {\cal {P}}\rangle  \neq 0
\end{equation}
Any operator ${\cal O}$  whose three point function $\langle TT{\cal O} \rangle $ is non-zero
will appear in the OPE of the stress tensor. In fact they will be more relevant than the 
stress tensor  if the 
operators have dimension $\Delta<4$,  and as relevant if the operator has dimension $\Delta =4$, 
see \cite{Osborn:1993cr} below equation (6.47). Thus even in the free field limit the operators 
considered can arise in the OPE of the stress tensor. 

Now we will show that the expectation values of the operators ${\cal O}_1$ and ${\cal O}_2$
is responsible for the modification of the sum rule. 
For this we  evaluate the following
\begin{eqnarray}
\label{sqexp}
 \langle {\cal O}_1 \rangle +  \langle {\cal O}_2\rangle &=& 
\frac{1}{N} ( \langle O_1\rangle ^2 + \langle O_2\rangle ^2 ) \\ \nonumber
&=& \frac{N^3 T_0^4}{192} \left\{ ( k_1 - k_2)^2 + ( k_1- k_3)^2 + ( k_2- k_3)^2 \right\}.
\end{eqnarray}
The first equality is due to large $N$ factorization. 
 In the second line we have substituted the expectation values form (\ref{expect}) and 
 re-expressed $r_+$ in terms of $T_0$. 
Note that  the 
 final line of (\ref{sqexp})  is proportional to  the additional term in the sum rule (\ref{sumcrule})
 with the exception of the factor of $N$. This fact is because  the structure 
 constant $B_{\mu\nu\rho}^a$ for these operators scales as $\frac{1}{N}$ 
 in the large $N$ limit as can be seen
 from the normalization of our operators. This additional factor of $\frac{1}{N}$ together
 with the $N^3$ in (\ref{sqexp}) gives the required $N^2$ scaling seen in the sum rule.  
Therefore we conclude these  additional terms in the shear sum rule
are due to the fact operators proportional to $O_i^2$ 
of dimension $4$ in the OPE which acquire 
non-trivial expectation values in the presence of chemical potentials.

\section{Shear sum rule for R-charged M2 and M5-branes}

We will now repeat the analysis and derive the shear sum rules for R-charged M2 and M5-branes. 
One of our motivations to do this is to verify if the modifications to the sum rules can be explained
using the same reasoning employed for the D3-brane case. 
Indeed we will find that that additional terms in the shear sum rule for the M2-brane can be 
explained in terms of expectation values of operators in the OPE of the stress tensor. 
For the M5-brane case we do not find any additional terms in the sum rule and this is 
consistent  with the fact that there are no operators of the requisite dimension in  the 
supergravity which acquires an expectation value.

\subsection{M2-branes}

We begin with the analysis of the M2-brane at finite chemical potential. 
In general there are  4 R-charges 
 corresponding to the Cartan's of $SO(8)$  therefore 
it is possible to turn on 4 independent chemical potentials.
The gravity dual of this system is given by the non-extremal  R-charged black hole in 
$AdS_4$
\cite{Duff:1999rk}. 
Using the same methods developed for the D3-brane case it is possible to show that the 
retarded Green's function of the $T_{xy}$ component of the stress tensor 
satisfies property 1. 
To ensure that it satisfies property 2, we need to subtract the divergent terms at
large frequencies. 
The constant term which arises in the Green's function as $\omega\rightarrow\infty$ essentially 
is the LHS of the sum rule. 
Therefore we will directly go ahead and evaluate the divergent pieces 
as $\omega\rightarrow \infty$ and isolate the term  which contributes to the sum rule.

The metric and the gauge field for the R-charged M2-brane with all the four charged turned on 
 is given by 
\begin{eqnarray}
\label{ch4met}
& & ds_4^2
  = \frac{16(\pi T_0 L)^2}{9u^2}{\cal H}^{1/2}
    \left( - \frac{f}{{\cal H}} dt^2 + dx^2 + dz^2 \right)
  + \frac{L^2}{f u^2}{\cal H}^{1/2}~du^2~, \\ \nonumber
& & A_{t}^{i}  = \frac{4}{3} \pi T_0 \sqrt{2 k_i\prod\limits_{i=1}^4
 (1+k_i)}~\frac{u}{H_i}~,~~~~~~
u=\f{r_+}{r},\qquad H_i = 1 + k_i u~,  \\ \nonumber
& &{\cal H}=\prod\limits_{i=1}^4 H_i,~~~~
f = {\cal H}-\prod\limits_{i=1}^4
 (1+k_i)u^{3} .
\end{eqnarray}
The scalars are given by 
\begin{equation}
\label{valscalm2}
X^i = {{\cal H}^{1/4}\over H_i(u)}. 
\end{equation}
The four scalars are not independent and constrained by $X_1X_2X_3X_4=1$.  
The above background is the solution of the equation of motion of the following
action 
\begin{eqnarray}
S &=& \frac{N^{3/2}\sqrt{2}}{24\pi L^2} \int d^4 x\sqrt{-g} {\cal L}, \\ \nonumber
{\cal L} &=& R - \frac{1}{2}( \partial \vec\phi)^2 + V(\phi)  - 
\frac{1}{4} \sum_{i=1}^4 e^{\vec a_i \cdot \vec \phi} (F^i)^2, 
\end{eqnarray}
where the fields four fields $X_i$ are related to the three independent fields 
$\vec \phi_i = ( \phi_1, \phi_2, \phi_3) $  by 
\begin{eqnarray}
\label{indpsc4}
& &  X_i = \exp( - \frac{1}{2} \vec a_i \cdot \vec  \phi ) , \\ \nonumber
& & \vec a_1 = ( 1, 1, 1), \quad \vec a_2 = ( 1, -1 , -1) , \quad  \vec a_3 = ( -1, 1, -1) , \vec a_4 = ( -1, -1, 1). 
\end{eqnarray}
The scalar potential is given by 
\bea
\label{sc4pot}
V(\phi)= \f{2}{L^2}(\cosh \phi_1 + \cosh \phi_2 + \cosh \phi_3).
\eea
Various thermodynamic quantities are summarized as follows: 
let us first  define 
\begin{equation}
  T_0=\f{3r_+}{4\pi L^2 }. 
\end{equation}
The energy density and pressure are  given by 
\begin{eqnarray}
   \epsilon
  &=& \sqrt{2}\, \pi^2\, \left( \frac{2}{3} \right)^4\,
    N^{3/2}\, T_0^3~\prod\limits_{i=1}^4(1 + k_i), \\ \nonumber
 P &=& \frac{\sqrt{2}\, \pi^2}{3}\, \left( \frac{2}{3} \right)^3\,
    N^{3/2}\, T_0^3~\prod\limits_{i=1}^4(1 + k_i)~.
\end{eqnarray}
The expression for entropy density and temperature  is given by  
\begin{eqnarray}
\label{temp4}
 s  &=&\sqrt{2}\pi^2 \left( \frac{2}{3} \right)^3 
N^{3/2}\, T_0^2~\prod\limits_{i=1}^4\sqrt{1 + k_i}, \\ \nonumber
T &=& \frac{T_0\left(3+\sum\limits_{j=1}^4 k_i+
\sum\limits_{j>i,i,j=1}^4k_i k_j-\prod\limits_{i=1}^4k_i \right)}{3\sqrt{\prod\limits_{i=1}^4(1 + k_i)}}.
\end{eqnarray}
The charge density and the chemical potential are given by 
\begin{eqnarray}
\label{chem4}
 \rho_i
  &=& \sqrt{2}\, \pi\, \left( \frac{1}{3} \right)^3\,
    N^{3/2}\, T_0^2~\sqrt{2\, k_i\prod\limits_{j=1}^4(1 + k_j) }, \\ \nonumber
\mu_i &=& \frac{4 \pi\, T_0}{3}\, \frac{1}{1+k_i}\sqrt{ 2\, k_i \prod\limits_{i=1}^4(1 + k_i)}.    \nn
\end{eqnarray}

To obtain the behaviour of the Green's function as $\omega\rightarrow\infty$,  we first 
write the equation of the minimally coupled scalar in the background given by (\ref{ch4met}) in 
terms of Fefferman-Graham coordinates. 
The metric given in (\ref{ch4met}) admits the following form 
in Fefferman-Graham coordinates.
\begin{equation}
\label{fgad4}
 ds^2 = \frac{1}{z^2} \left( ( -1 + \sum_{i=1}^\infty a_i z^{i} ) dt^2 
+ ( 1 + \sum_{i=1}^\infty b_i  z^{i} ) ( d\vec x^2 ) + dz^2 \right). 
\end{equation}
Here we are working with the scaled variables 
\bea
z=\f{r_+}{r},\qquad i\l=\f{L^2}{r_+}\o. 
\eea
In the   Fefferman-Graham expansion given in (\ref{fgad4})
it can be shown that  the terms  $a_1, b_1$ are zero. 
As we will see subsequently, the asymptotic behaviour of the 
retarded Green's function is essentially determined   by the coefficients
$a_2,b_2, a_3,b_3$ which are evaluated in (\ref{val4ab}). 
Now the equation for the minimally coupled scalar in this metric 
(\ref{fgad4}) is given by
\begin{equation}
 \phi'' -\frac{1}{z} \left( 2 + \sum_{i=1}^\infty \tilde a_i (z)^i \right) \phi'
 - \lambda^2 ( 1 + \sum_{i =1}^\infty \tilde b_i (z)^i ) \phi =0, 
\end{equation}
 where $\tilde a_i, \tilde b_i$ can be obtained 
given $a_i, b_i$.  For the coefficients of interest,  this relation is given by 
\bea
 \tilde a_2 =  ( a_2 -2b_2) , \qquad \tilde b_2 = a_2,\nn\\
  \tilde a_3 = \f{3}{2} ( a_3 -2b_3) , \qquad \tilde b_3 = a_3. 
\eea
We  now rescale  the coordinates  by defining 
\begin{equation}
 y = \lambda z. 
\end{equation}
Then the equation for the minimally coupled scalar reduces to 
\begin{equation}
\label{scaleq4}
 \phi''(y) - \frac{1}{y} \left( 2 + \sum_{i=1}^\infty \tilde 
a_i (\frac{y}{\lambda}) ^i \right) \phi'(y) - 
( 1 + \sum_{i =1}^\infty \tilde b_i ( \frac{y}{\lambda})^i ) \phi( y) =0. 
\end{equation}
Note that at $\lambda$ strictly infinity, the equation 
reduces to that of the minimally coupled scalar in 
the $AdS_4$ or the $T=0$ background. 
To solve the equation perturbatively in $\f{1}{\l}$ we define the quantity 
\be
g(y) =\f{\phi'}{\phi}. 
\ee
From the expansion of the equation in (\ref{scaleq4}), we see that we can expand g as 
\bea
g=g_0+\f{1}{\l^2}g_1+\f{1}{\l^3}g_2+\ldots. 
\eea
The equations satisfied by these can be obtain from (\ref{scaleq4}) and the first few equations are 
\bea
\label{expad4}
& & g_0'+g_0^2-\f{2}{y}g_0-1=0,\nn\\
& & g_1'+2(g_0-\f{1}{y})g_1-\tilde{a_2}y g_0-\tilde{b_2}y^2=0,\nn\\
& & g_2'+2(g_0-\f{1}{y})g_2-\tilde{a_3}y^2 g_0-\tilde{b_3}y^3=0.
\eea
The two independent solutions for the first equation in (\ref{expad4}) are 
\bea
g_0^{(1)} = -\frac{K_{1/2}(y)}{K_{3/2}(y) }= -\f{y}{1+y}, \qquad 
g_0^{(2)} =   \frac{I_{1/2}(y)}{I_{3/2}(y)}= \frac{ x \sinh x}{ x \cosh x -\sinh x}.  
\eea
As in the case of the D3-branes we can solve for the two independent solutions order by order in 
$\frac{1}{\lambda}$ around these solutions. 
Just as in the case of  the D3-brane situation we find that  the solution around $g_0^{(1)}$ determines
the leading asymptotic behaviour of the Green's function as $\lambda \rightarrow \infty$. 
This is because of the condition that at $\lambda\rightarrow\infty$, the solution must reduce to that
of the $T=0$ case.  $I_{3/2}(y)$ is a growing solution as $y\rightarrow\infty$ and thus is 
not allowed when $T=0$. 
The leading corrections around the solution $g_0^{(0)}$ is given by 
\bea
g_1 &=&-y^2(\f{a_2}{2}+b_2\f{2y+3}{2(1+y)^2}),\nn\\
g_2 &=& \f{3}{4}(a_3-2b_3)y^2-\f{a_3 y^2}{8(1+y)^2}(9+18y+14y^2+4y^3).
\eea
The retarded Green's function is then given by 
\begin{eqnarray}
\label{gm2}
\lim_{\omega\rightarrow\infty}  G_R(\omega, T) &=& 
\lim_{\omega\rightarrow\infty} \frac{ \sqrt{2} N^{3/2} r_+^3}{24\pi L^6} \lim_{y\rightarrow 0}
 \frac{1 }{y^2} \left( \lambda^3 g_0^{(1)}(y)  +  \lambda g_1^{(1)}(y) + 
g_2^{(1)}(y)  \right) \nonumber \\ 
&& + G_{\rm{contact}}  + G_{\rm{counter}}. 
\end{eqnarray}
Substituting the form for $g_1^{(1)}$ and $g_2^{(1)}$ in the limit we have
\begin{eqnarray}
\label{limg1g2}
\lambda \lim_{y\rightarrow 0} \frac{g_1^{(1)}(y)}{y^2}  = - \frac{\lambda}{2} \left( a_2 + 3 b_2\right),  
\qquad 
 \lim_{y\rightarrow 0} \frac{g_2^{(1)}(y)}{y^2} = - \frac{3}{8} ( a_3 + 4b_3). 
\end{eqnarray}
From the expressions in   (\ref{limg1g2}),  it is clear that there are two sources of 
divergence.  The term  proportional to $\lambda^3$ which is identical to the contribution of 
the $T=0$ Green's function. The second divergence is due to the 
term proportional to $\lambda$.  This gives a linear divergence in $\omega$ and we need to 
subtract this also along with  the constant term from the limit of $g_2^{(1)}$  to ensure that 
property 2. is satisfied.
Therefore we  define
\begin{eqnarray}
 \delta G_R(\omega, T) &=& G_R(\omega, T) - G_R(\omega, 0) - G_{\rm{contact}} \\ \nonumber
\;\;& &  +i \delta \rho(\omega)
 + \frac{\sqrt{2}N^{3/2} r_+^3}{24\pi L
^6} \frac{3}{8} ( a_3 +4 b_3). 
\end{eqnarray}
where
\begin{eqnarray}
\label{deltarho}
& & \delta\rho (\omega)  =   -\frac{\sqrt{2}N^{3/2} r_+^2\omega }{24\pi L^4} \frac{1}{2}( a_2 + 3b_2), 
\\ \nonumber
& =& \frac{\sqrt{2} \pi T_0^2 N^{3/2} \omega}{432}  \left( 
( k_1 + k_2 - k_3 -k_4)^2 + ( k_1 -k_2+k_3 -k_4)^2 + ( k_1-k_2-k_3+k_4)^2 
\right). 
\end{eqnarray}
Note that we have subtracted the required terms to ensure that $\delta G_R(\omega, T)$ 
obeys property 2 to apply Cauchy's theorem. 
The RHS  of the sum rule therefore involves the following integrand
\begin{eqnarray}
{\rm{Im}}\,  \delta G_R(\omega, T) &=& \rho(\omega) + \delta\rho(\omega) - \rho(\omega, 0) , 
\\ \nonumber
&=& \tilde \rho(\omega, T) - \rho(\omega, 0). 
\end{eqnarray}
The LHS of the sum rule is obtained by evaluating 
\begin{eqnarray}
\delta G_R(0, T) &=&   \frac{\sqrt{2}N^{3/2} r_+^3}{24\pi L^6} \frac{3}{8} ( a_3 + b_3), \\ \nonumber
&=& \f{3}{8}\epsilon +\frac{\sqrt{2}\pi N^{3/2} T_0^3}{216} (k_1 + k_2 - k_3 - k_4) (k_1 - k_2 + k_3 - k_4) (k_1 - k_2 - k_3 + k_4). 
\end{eqnarray}
Here we have substituted the values of $a_3, a_4$ from (\ref{val4ab}). 
Thus the sum rule for the M2-brane theory in the presence of 
chemical potential is given by 
\begin{eqnarray}
\label{lhsm2}
 && \frac{3}{8} \epsilon +\f{\sqrt{2}\pi^2 N^{3/2} T_{0}^{3}}{216 } 
( k_1 + k_2 - k_3-k_4) ( k_1 - k_2 + k_3 - k_4) ( k_1 -k_2-k_3+k_4) 
 \nonumber \\ 
 & &= \frac{1}{\pi} \int_{-\infty}^\infty  \frac{d\omega}{\omega}
\left( \tilde \rho(\omega) - \rho_{T=0} (\omega) \right). 
\end{eqnarray}
Note that $\tilde \rho(\omega)$ is defined by 
\begin{equation}
\label{tilrho}
\tilde \rho(\omega)_T = 
\rho(\omega)_T + \delta\rho(\omega), 
\end{equation}
where $\delta\rho(\omega)$ is given in  (\ref{deltarho}).

From the above analysis we see that the sum rule is essentially due to the 
high frequency behaviour of the Green's function.  
High frequency behaviour of the Green's function can be extracted 
from the OPE of the stress tensor. We will now show that the linear divergence in 
$\omega$ and the additional term involving the chemical potentials in the LHS of the sum 
rule are due to the presence of certain operators in the OPE which 
get expectation values in the presence of the sum rule. 
The general form for the OPE of the stress tensor in this case is given by 
\be
\label{ope4}
 T_{\mu\nu}( x) T_{\rho\sigma}(0) \sim C_T \frac{ I_{\mu\nu, \rho\sigma} }{x^6} + 
\hat A_{\mu\nu\rho\sigma\alpha\beta}(x) T_{\alpha\beta}(0) +
B_{\mu\nu\rho\sigma}^a (x) {\cal O}_a (0). 
\ee
Here we have included additional terms in the OPE which arise if there are scalars denoted by 
$ {\cal O}_a $ are present in the theory and the three point function$ 
 \langle T_{\mu\nu}T_{\rho\sigma} {\cal O}_a\rangle$ is non-zero.
Note that now the stress tensor is a dimension 3 operator in this case.
From a similar analysis as in the case of the D3-brane case, we see that a linear
divergence  can be explained due the presence of 
operators ${\cal O }$of dimension 1 in the OPE. 
The constant terms must arise from operators of dimension 3. 
Since the expectation value of the stress tensor accounts for the term
$\frac{3}{8} \epsilon$ in the sum rule, the additional terms in the LHS of the sum rule 
must arise from other operators ${\cal O}_a$ of dimension 3. 
Note that the $\omega$ divergence in (\ref{deltarho}) and the additional 
term  present in the LHS of the sum rule in (\ref{lhsm2})  vanish when all the chemical potentials
are equal. In such a situation from the solution  given in (\ref{ch4met}) one sees that
there is non-trivial  gauge field but the scalars given in (\ref{valscalm2})  become trivial. 
Thus the presence of expectation values of these operators must be 
due to the scalars. 

Let us now examine the scalars in detail. 
Expanding the scalar potential given in (\ref{sc4pot}) to quadratic order we find that 
all the three scalars $\phi_i$ are of equal mass with the mass given by 
\begin{equation}
m^2 L^2 = -2. 
\end{equation}
The scaling dimension of the operator dual to the scalars can be found by the mass-dimension 
relation 
\begin{equation}
\Delta(\Delta -3) = m^2 L^2. 
\end{equation}
From this equation we see that the dimension of the operators $\Delta$ can be chosen to be 
$1$ or $2$. When $\Delta$ is chosen to be $2$ the quantization of the scalar is conventionally
known as the standard quantization. One the other hand when $\Delta$ is chosen to 
be $1$ the quantization of the scalar obey the alternate quantization. 
For the neutral 
scalars in the M2-brane background it has been seen earlier 
\cite{Bobev:2009ms,Donos:2011ut} that one must chose the 
alternate quantization. 
Let us denote the three operators dual to these bulk scalar fields as $O_i$ which have
dimension $1$ each. 
We now read out the expectation values of these operators in the non-extremal M2-brane
background following \cite{Marolf:2006nd}. 
The expansion of $\phi_1$ close to the boundary is of the form 
\begin{equation}
 \phi_{i}= \alpha_i \f{1}{r}+ L^2 \beta_i \f{1}{r^2} + \cdots . 
\end{equation}
The factor of $L$ has been re-instated so that $\beta$ has the appropriate dimensions. 
For the alternate quantization, the expectation value of the operators is given by the coefficient 
$\frac{N^{3/2} \sqrt{2}}{24\pi L^2} \alpha_i$. Expanding the scalar fields  $\phi_i$ 
defined in (\ref{indpsc4}) near the boundary
and using  (\ref{valscalm2}) 
 we obtain
 \begin{eqnarray}
 \langle O_1 \rangle
 &=& -\left( \frac{N^{\frac{3}{2}}\sqrt{2}}{24\pi}\right)
  \frac{r_+}{2 L^2}(k_1+k_2-k_3-k_4),\nonumber\\
\langle O_2 \rangle
&=& -\left( \frac{N^{\frac{3}{2}}\sqrt{2}}{24\pi }\right)
\frac{r_+}{2 L^2}(k_1-k_2+k_3-k_4),\nonumber\\
 \langle O_3 \rangle 
 &=& -\left( \frac{N^{\frac{3}{2}}\sqrt{2}}{24\pi}\right)
\frac{r_+}{2 L^2}(k_1-k_2-k_3+k_4).
\end{eqnarray}
From these expectation values we see that the candidate operators which are responsible for the 
linear $\omega$ divergence given in (\ref{deltarho}) are proportional  $O_i^2$.
Indeed taking the sum or squares of these operators we obtain
\begin{eqnarray}
\label{sumsq}
& & \sum_{i=1}^2\langle O_i^2 \rangle = \sum_{i=1}^3 \langle O_i \rangle^2 , \\ \nonumber
 & & = \frac{N^3}{648} \left( 
 (k_1+k_2-k_3-k_4)^2 + (k_1-k_2+k_3-k_4)^2 + (k_1-k_2-k_3+k_4)^2 \right) . \\ \nonumber
 \end{eqnarray}
 In the first line we have assumed large $N$ factorization. 
 Note that the last line in (\ref{sumsq}) is proportional to the linear omega divergence 
 seen in (\ref{deltarho}). The factors of $N$ will present in (\ref{sumsq}) will agree 
 with that present in (\ref{deltarho}) if one assumes the structure constant $B_{\mu\nu\rho}^a$
with the normalization of the operators used scales as $N^{-3/2}$. Indeed 
this is what is seen from the gravity analysis\footnote{Since we do 
not have an explicit theory for the M2-branes at present we cannot understand
the $N$ scalings from a field theory analysis for the M2-branes.}. 
 We can also obtain the additional term present in the LHS of the sum rule given in (\ref{lhsm2}). 
 Note that for this we require an operator of dimension 3. We see that the required operator
 is proportional to $O_1O_2O_3$. Indeed we find 
 \begin{eqnarray}
& &  \langle O_1 O_2 O_2 \rangle = 
 \langle O_1 \rangle \langle O_2 \rangle \langle O_2 \rangle,  \\ \nonumber
& & =\frac{N^{\frac{9}{2} }\sqrt{2}}{2^5 3^6}
(k_1+k_2-k_3-k_4)(k_1-k_2+k_3-k_4)(k_1-k_2-k_3+k_4). 
\end{eqnarray}
which proportional to  the additional term in (\ref{lhsm2}) with the exception of the $N$ scaling. 
The $N$ scaling can be understood if one assumes that the structure constant 
$B_{\mu\nu\rho\sigma}^a$ scales as $N^{-3}$ for these operators. 
Thus the additional term in the 
LHS of the sum rule (\ref{lhsm2}) is due to these additional operators in the OPE which 
gain expectation value in presence of the chemical potential. 

\subsection{M5-branes }

In this section we examine the dual of M5-brane at finite chemical potential and finite temperature and re-derive the shear sum rule. 
Since there are 2 R-charges corresponding to the Cartan's of SO(5) it is possible to turn on  2 independent chemical potential,since there are no constrain equation relating this two R-charges.
Using the differential equation of the massless minimally coupled scalar in this background we obtain the retarded Green's function of the $T_{xy}$ component of the stress tensor.
Examining this differential equation and the 
same method employed for he D3-brane case it can be shown that the regulated Green's function satisfies both property 1 and property 2 which are necessary for deriving the sum rule.
However  as will see  the shear sum rule for the 
case of M5-brane is not modified. We argue that this 
 must be the case since there are no scalars in the M5-brane gravity background which 
have the appropriate dimensions to modify the OPE of the stress tensor. 

The metric and the gauge field for the R-charged M5-brane with all the charges turned 
on  is given by \cite{Duff:1999rk}
\begin{eqnarray}
\label{ads7}
&&ds_7^2
  = \frac{4(\pi T_0 L)^2}{9u}{\cal H}^{1/5}
    \left( - \frac{f}{{\cal H}} dt^2 + dx_1^2 + \cdots + dx_4^2 + dz^2 \right)
  + \frac{L^2}{4 f u^2}{\cal H}^{1/5}~du^2~,  \nonumber \\ 
&&A_{t}  = \frac{2}{3} \pi T_0 \sqrt{2 k_i\prod\limits_{i=1}^2(1+k_i)}~\frac{u^2}{H_i}~,~~~~~H_i = 1 + k_i u^2~,\nn\\
&&T_0=\f{3 r_+}{2 \pi L^2},\qquad \epsilon=\f{5 \pi^3}{2}\f{2^7}{3^7}N^3 T_0^6\prod_{i=1}^{2}(1+k_i)
\\ \nonumber
&&H_i = 1 + k_i u^2~, ~~~~~~~~
{\cal H}=\prod\limits_{i=1}^2 H_i, \quad
f = {\cal H}-\prod\limits_{i=1}^2(1+k_i)u^{3}~. 
\end{eqnarray}
The background solution for the two scalars are given by 
\begin{equation}
\label{scalm5}
X^i = {{\cal H}^{2/5}\over H_i(u)} ,
\end{equation}
where $u=\frac{r_+^2}{r^2}$.
The above background is the solution of the equation of motion of the action is
\begin{eqnarray}
S &=& \frac{N^3}{6 \pi^3 L^5} \int d^7 x\sqrt{-g} {\cal L}, \\ \nonumber
{\cal L} &=& R - \frac{1}{2}( \partial \vec\phi)^2 + V(\phi)  - 
\frac{1}{4} \sum_{i=1}^2 e^{\vec a_i \cdot \vec \phi} (F^i)^2. 
\end{eqnarray}
Where the two  scalar fields $X_i's$ are related to $\vec{\phi}=(\phi_1,\phi_2)$ b;y 
\bea
&& X_i=e^{-\f{1}{2}\vec{a}\cdot \vec \phi },\nonumber\\
&& \vec{a_1} =(\sqrt{2},\sqrt{\f{2}{5}}),\qquad 
\vec{a_2}=(-\sqrt{2},\sqrt{\f{2}{5}}). 
\eea
Note that the two scalars $X_i$ are independent here, unlike the situation in 
the case of the D3-branes and M2-branes.
The  scalar potential  $V$ is given by 
\be
\label{m5pot}
V= \frac{4}{L^2} \left( -4X_1 X_2-2X_1^{-1}X_2^{-2}-2X_1^{-2}X_2^{-1}+\f{1}{2}(X_1 X_2)^{-4} \right).
\ee
The  leading terms in the large frequency expansion of the 
equation for $xy$ component of the metric field in the Fefferman-Graham
coordinate system is given by
\begin{equation}
\label{ads7min}
 \phi''-\f{1}{y}(5+\f{3(a_3-5b_3)y^6}{\l^6})\phi'-(1+a_3 \f{y^6}{\l^6})\phi=0.
\end{equation}
where we have defined
\begin{equation}
y = \lambda z = \lambda \frac{r_+}{r}, \qquad i \lambda = \frac{L^2}{r_+}\omega. 
\end{equation}
The values of $a_3, b_3$ are evaluated in  in (\ref{vala3b3m5}). 
Let us define
\begin{equation}
\tilde{a}_3=3(a_3-5b_3), \qquad  \tilde{b}_3=a_3. 
\end{equation}
Since the Green's function is proportional to 
$g=\f{\phi'}{\phi}$,  we can write the  equation in (\ref{ads7min}) as 
\begin{equation}
\label{ads7g}
 g'+g^2-\f{1}{y}(5+\f{\tilde{a}_3y^6}{\l^6})g-(1+\tilde{b}_3\f{y^6}{\l^6})=0.
\end{equation}
We then find the  solution to the leading order  by writing $g$ as 
\be
g=g_0+\f{1}{\l^6}g_1+ \cdots. 
\ee
The equations satisfied by each of the terms in the 
expansion of $g$  can be obtain from (\ref{ads7g}).  They are given by 
\bea
& & g_0'+g_0^2-\f{5}{y}g_0-1=0,\nn\\
& & g_1'+2g_0 g_1-\f{5}{y}g_1-\tilde{a}_3 y^5 g_0-\tilde{b}_3 y^6=0. 
\eea
The relevant solution for our analysis are 
\bea
g_0^{(1)}=-\f{K_2(y)}{K_3(y)},\qquad 
g_1^{(1)} = \f{\tilde{a}_3 y^5}{2}+\f{\tilde{b}_3 y^7}{14}(1-\f{K_4^2}{K_3^2}). 
\eea
The retarded Green's function is then given by 
\begin{eqnarray}
\label{gm3}
\lim_{\omega\rightarrow\infty}  G_R(\omega, T) &=& 
\lim_{\omega\rightarrow\infty} \frac{ N^{3} r_+^6}{6\pi^3 L^{12}} \lim_{y\rightarrow 0}
 \frac{1 }{y^2} \left( \lambda^6 g_0^{(1)}(y)  +   g_1^{(1)}(y)  
 \right) \nonumber \\ 
&& + G_{\rm{contact}}  + G_{\rm{counter}}. 
\end{eqnarray}
The term proportional to $\l^6$ which is identical to the contribution of the $T=0$ 
retarded Green's function. It  is divergent in the limit $\l\rightarrow \infty$.
To ensure that the property 2 is satisfied we need to subtract the divergent piece and the
constant pieces from the Green's function to write the regulated Green's function. 
Therefore we define 
\begin{eqnarray}
 \delta G_R(\omega, T) &=& G_R(\omega, T) - G_R(\omega, 0) - G_{\rm{contact}} \\ \nonumber
\;\;& &  
 +  \frac{N^3}{3\pi^3}\f{15r_+^6}{14 L^{12}}(7 b_3+a_3).
\end{eqnarray}
The LHS of the sum rule is obtain by evaluating
\begin{eqnarray}
\delta G_R(0,T) &=& \frac{N^3}{3\pi^3}\f{15}{7 }\left(\frac{2\pi T_0}{3}\right)^{6}(1+\kappa_1)(1+\kappa_2),  \nonumber\\
&=&\frac{3}{7}\epsilon.
\end{eqnarray}
Here we have substituted the values of $a_3,b_3$ from (\ref{vala3b3m5}).
Thus the sum rule for the M5-brane in the presence of chemical potential is given by
\begin{eqnarray}
 && \frac{3}{7} \epsilon  = \frac{1}{\pi} \int_{-\infty}^\infty  \frac{d\omega}{\omega}
\left( \rho(\omega) - \rho_{T=0} (\omega) \right). 
\end{eqnarray}

Note that for the   case  of R-charged M5-branes 
 there is no additional contribution in the shear sum rule
compared to uncharged case. This fact can be understood as follows. 
Let us examine the masses of the scalars in the theory. From expanding
the potential for the scalars in (\ref{m5pot}) we see that the two scalars $\phi_i$ in the 
theory have mass given by by 
\begin{equation}
m^2L^2 = - 8. 
\end{equation}
From the mass-dimension relation we see that there are two possible choices
of mass for the operators dual to these. They are $\Delta =2$ or $\Delta =4$. 
The  stress tensor for the 6-dimensional theory of the M5-brane
is an operator of dimension $6$. Thus if there are terms which are finite in the large frequency 
expansion of the OPE, they must be of dimension $6$.
The general form for the asymptotic expansion of the scalar field in this case
is given by 
\begin{equation}
\phi_i = \frac{\alpha_i}{r^2} + \frac{\beta_i}{r^4} + \cdots. 
\end{equation}
From examining the background value for the scalar field given in (\ref{scalm5}) we see that 
for both the scalar fields, $\alpha_i =0$ and $\beta_i\neq 0$. 
In the standard quantization of the scalar fields, 
the dual operators have dimension $\Delta =4$ and 
the expectation value of the scalar field is proportional to the value of $\beta_i$. But in this situation
there is no operator which has a dimension $\Delta =6$ 
 which gains expectation value. This implies there is no finite term in the large frequency expansion 
of the OPE of the stress tensor \footnote{The operator dual to the scalar field with dimension 
$\Delta =4$ can give rise to a $\omega^2$ divergence if there the 3 point function of this operator
with the stress tensor $\langle T_{\mu\nu} T_{\rho\sigma} {\cal O} \rangle$ is not zero.
From our gravity analysis since there is no divergence it is clear that this three point 
function vanishes.}. 
If we choose the alternate quantization, then the operators dual to the fields $\phi_i$ have 
dimension $2$, but from the expansion of the scalar field we see that their expectation value 
is proportional  to the value of $\alpha_i$ which is zero. Thus in the alternate quantization 
the operator dual to the scalar field does not gain expectation value. Therefore it cannot contribute 
to either a divergence or the finite term in the OPE of the stress tensor. 
These arguments imply that there is there is no correction to the sum rule which is consistent 
with our explicit calculation.

\section{Conclusions}

By examining the shear sum rule for the case of
R-charged D3-branes  and M2-branes  we arrive at the general observation that sum rule acquires additional terms 
proportional to the chemical potential. These additional terms are the result of 
expectation values of operators of dimension $4$ and dimension $3$ for the case 
of D3-branes and M2-branes respectively. 
For the case of the M5-brane this no change in the sum rule at finite chemical potential. 
This is explained by the fact that there is no operator of dimension $6$ whose expectation value
is turned on at finite chemical potential. 
Our analysis indicates that the  LHS of the sum rule contains information of the OPE coefficients of the 
stress tensor with scalar operators of the  appropriate dimensions. Thus this approach can be
used to compute three point of functions of the stress tensor with these operators from 
gravity.  This will involve writing down the 
relating  the coefficient $B_{\mu\nu\rho\sigma }^a$ given in (\ref{opeop})  in terms 
of the coefficient which occurs  in the three point function using a similar analysis of
\cite{Osborn:1993cr}. It will be interesting to explicitly check the three point function 
obtained in this manner  against a direct evaluation 
three point functions from  gravity.  

Another direction to extend this work would be to examine sum rules for the R-charge correlators 
in ${\cal N}=4$ Yang-Mills at finite chemical potential. It will be interesting to see if the  effects 
of triangle anomalies can be seen from the sum rule just as hydrodynamics  was modified 
on the account of triangle anomalies \cite{Son:2009tf}. 
Finally we mention that chemical potential is an  important parameter in the 
QCD phase diagram and it will be interesting to  see if sum rules can be derived 
in this context.  

\acknowledgments

We  wish to thank  Avinash Dhar,  Rajesh Gopakumar, Kavita Jain,
Gautam Mandal,  Shiraz Minwalla, Ashoke Sen and Dam  Son for useful discussions. 
S. J would like to thank CHEP, IISc, Bangalore  for hospitality where part of this work was done. 
The work of J.R.D  is partially supported by the Ramanujan fellowship DST-SR/S2/RJN-59/2009.

\appendix

\section{ Fefferman-Graham coordinates}

In this appendix we will detail the co-ordinate transformation which takes the 
metrics discussed in this paper to the Fefferman-Graham coordinates. 
We will see that for the non-extremal R-charged D3 and  M5-branes, the metric in
 Fefferman-Graham co-ordinates takes the form
\begin{equation}
 ds^2 = \frac{1}{z^{\prime 2} \tilde L^{2}} \left( ( -1 + \sum_{i=1}^\infty a_i (z^{\prime 2})^{i} ) dt^2 
+ ( 1 + \sum_{i=1}^\infty b_i ( z^{\prime 2})^{i} ) ( d\vec x^2 ) + dz^{\prime 2} \right).  
\end{equation}
For the case of the non-extremal R-charged M2-brane the metric in
 Fefferman-Graham  co-ordinates takes the form 
\begin{equation}
 ds^2 = \frac{1}{z^{\prime 2} \tilde L^{2}} \left( ( -1 + \sum_{i=1}^\infty a_i (z^{\prime} )^{i} ) dt^2 
+ ( 1 + \sum_{i=1}^\infty b_i ( z^\prime )^{i} ) ( d\vec x^2 ) + dz^{\prime 2} \right) , 
\end{equation}
here $\tilde L = L/r_+$. 
In this appendix we will evaluate the coefficients $a_i, b_i$ which are necessary to obtain 
the sum rule. 

\subsubsection*{Non-extremal R-charged D3-brane}

The metric R-charged D3-brane metric given in 
in (\ref{chgmet}) can be transformed to the Fefferman-Graham form by first  redefining the 
co-ordinate $z = \frac{r_+}{r}$ in terms of the Fefferman-Graham co-ordinate $z'$ as 
a power series 
\bea
\label{cortran}
z &=& \Omega({z'}),\\ \nonumber
&=& {z^\prime}(1+\a z^{\prime 2} +\b z^{\prime 4} +O(z^{\prime 6} )). 
\eea
After expanding the coefficients of the metric (\ref{chgmet}) as a power series in $z$ around
$z=0$  we substitute the expansion (\ref{cortran}) in the metric (\ref{chgmet}).
We then determine the 
coefficients $\alpha, \beta, \cdots$ by requiring that the coefficient of the 
$dz^{\prime 2}$ term reduces to $\frac{1}{z^{\prime 2}}$. 
Performing this procedure to $O(z^2)$ results in the following values of $\alpha, \beta$
\bea
\a&=&\f{1}{6 }(k_1+k_2+k_3), \\ \nonumber
\b&=&\f{1}{72 }\lbrace(k_1^2+k_2^2+k_3^2)-(k_1k_2+k_2k_3+k_3k_1)-9(1+k_1+k_2+k_3+k_1k_2k_3)\rbrace. 
\eea
Finally we substitute these values back in the remaining coefficients of the metric to obtain
\bea
\label{gnfg}
a_1 &=& 0, \qquad b_1 =0, \\ \nonumber
a_2 &=&\frac{1}{36}\lbrace 27(1+ k_1 +k_2+k_3+k_1 k_2 k_3)+25(k_1 k_2+k_2 k_3+k_3 k_1) \\ \nonumber
&+& 2(k_1^2+k_2^2+k_3^2)\rbrace, \\ \nonumber
b_2&=&\frac{1}{36}\lbrace 9(1+ k_1 +k_2+k_3+k_1 k_2 k_3)+11(k_1 k_2+k_2 k_3+k_3 k_1) \\ \nonumber
&-& 2(k_1^2+k_2^2+k_3^2)\rbrace. 
\eea

\subsubsection*{Non-extremal R-charged M2-brane}

We now perform the same procedure for the metric given in (\ref{ch4met}). 
Let us define the co-ordinate $z$ as 
\bea
z &=& \Omega(\bar{z^\prime }),\\ \nonumber
&=& {z^\prime }(1+\a {z}^{\prime }+\b {z}^{\prime 2}+\g {z}^{\prime 3}+O({z}^{\prime 4} )). 
\eea
Demanding that the coefficient of $d{z}^{\prime 2}$ to be 
of the form of $\f{1}{z^{\prime 2}}$ to $ O( z^{\prime 4}) $ in the new coordinate system 
determines $\alpha, \beta, \gamma$. They are given by 
\bea
\a &=& \f{1}{4} (k_1 + k_2 + k_3 + k_4),\\ \nonumber
\b &=& \f{1}{64} \lbrace(k_{1}^{2}+k_{2}^{2}+k_{3}^{2}+k_{4}^{2})+10( k_1 k_2+ k_1 k_3+ k_1 k_4+ k_2 k_3+ k_2 k_4+ k_3 k_4)\rbrace, \\ \nonumber  
\g &=& \f{1}{384}\lbrace(k_1^3+k_2^3+k_3^3+k_4^3)\\ \nonumber &+&
11(k_1^2k_2+k_2^2k_1+k_1^2k_3+k_3^2k_1+k_1^2k_4+k_4^2k_1+k_2^2k_3+k_3^2k_2+k_2^2k_4+k_4^2k_2+k_3^2k_4+k_4^2k_3) \\ \nonumber
&-&2(k_1k_2k_3+k_1k_2k_4+k_1k_3k_4+k_2k_3k_4)-64(1+k_1+k_2+k_3+k_4 \\ \nonumber
&+&k_1k_2+k_1k_3+k_1k_4+k_2k_3+k_2k_4+k_3k_4+k_1k_2k_3k_4)
\rbrace. 
\eea
We then substitute these values back in the other coefficients of the metric to 
determine the values of $a_i, b_i$ which results in the following values. 
\bea
\label{val4ab}
a_1&=&b_1 =0, \\ \nonumber
a_2 &=& \f{1}{32}\lbrace 3( k_{1}^{2}+k_{2}^{2}+k_{3}^{2}+k_{4}^{2})-2(k_1k_2+k_1k_3+k_1k_4+k_2k_3+k_2k_4+k_3k_4)
\rbrace,\\ \nonumber
a_3 &=&\f{1}{24}\lbrace -(k_1^3+k_2^3+k_3^3+k_4^3)+14(k_1k_2k_3+k_1k_2k_4+k_1k_3k_4+k_2k_3k_4) \\ \nonumber
&+&(k_1^2k_2+k_2^2k_1+k_1^2k_3+k_3^2k_1+k_1^2k_4+k_4^2k_1+k_2^2k_3+k_3^2k_2+k_2^2k_4+k_4^2k_2+k_3^2k_4+k_4^2k_3) \\ \nonumber
&+&16(1+k_1+k_2+k_3+k_4+k_1k_2+k_1k_3+k_1k_4+k_2k_3+k_2k_4+k_3k_4+k_1k_2k_3k_4)
\rbrace, \\ \nonumber
b_ 2&=& -\f{1}{32}\lbrace 3( k_{1}^{2}+k_{2}^{2}+k_{3}^{2}+k_{4}^{2})-2(k_1k_2+k_1k_3+k_1k_4+k_2k_3+k_2k_4+k_3k_4)
\rbrace, \\ \nonumber
b_3 &=&\f{1}{24}\lbrace (k_1^3+k_2^3+k_3^3+k_4^3)+10(k_1k_2k_3+k_1k_2k_4+k_1k_3k_4+k_2k_3k_4) \\ \nonumber
&-&(k_1^2k_2+k_2^2k_1+k_1^2k_3+k_3^2k_1+k_1^2k_4+k_4^2k_1+k_2^2k_3+k_3^2k_2+k_2^2k_4+k_4^2k_2+k_3^2k_4+k_4^2k_3)
 \\ \nonumber
&+&8(1+k_1+k_2+k_3+k_4+k_1k_2+k_1k_3+k_1k_4+k_2k_3+k_2k_4+k_3k_4+k_1k_2k_3k_4)\rbrace.
\eea

\subsubsection*{ Non-extremal R-charged M5-brane}

We follow the same procedure again. 
We first expand the metric around $z=0$, retaining terms to order ${\cal{O}}(z^6)$ and then apply the coordinate transformation
\bea
z &=& \Omega({z^\prime }),\nn\\
&=& {z^\prime }(1+\a {z}^{\prime 2}+\b {z}^{\prime 4}+\g \bar{z}^{6}+O({z}^{\prime 7})). 
\eea
By  demanding that the coefficient of $d{z}^{\prime 2}$ is of the form of $\f{1}{ z ^{\prime 2} }$ 
to $O ({z}^{\prime 7} )$ in the new coordinate system we can determine $\a$,$\b$ and $\g$ .
They are given by 
\bea
\a = 0,\qquad 
\b = \f{k_1+ k_2}{10},\qquad 
\g = -\f{(1+k_1)(1+k_2)}{12}. 
\eea
Substituting these  values back 
in the metric we determine the other coefficients of the metric to be given by 
\bea
\label{vala3b3m5}
a_1 =  b_1=0, & \qquad&  
a_2 =  b_2=0, \\ \nonumber
a_3 = \f{5}{6}(1+k_1)(1+k_2),& \qquad&  b_3=\f{1}{6}(1+k_1)(1+k_2). 
\eea

\providecommand{\href}[2]{#2}\begingroup\raggedright\endgroup

\end{document}